\documentclass{aa1}

\usepackage{graphics}

\newcommand\citetb[1]{[\cite{#1}]}

\begin{document}

   \thesaurus{20            
             (04.03.1;)     
             (08.02.5;)     
             (08.06.3;)}    

\title{A catalogue of symbiotic stars}

\author{K.\ Belczy{\'n}ski \inst{1} \and J.\ Miko{\l}ajewska \inst{1} 
        \and U.\ Munari \inst{2} \and  R.\ J.\ Ivison \inst{3}
         \and M.\ Friedjung \inst{4} }

\offprints{K.\ Belczy{\'n}ski}
   \mail{kabel@camk.edu.pl}

\institute{
 Nicolaus Copernicus Astronomical Center,
 Bartycka 18, 00-716 Warszawa, Poland
\and
 Osservatorio Astronomico di Padova, Sede di Asiago, 
 I-36012 Asiago (Vicenza), Italy
\and
 Dept of Physics \& Astronomy, University College London,
 Gower Street, London WC1E 6BT, England
\and
 Institute d'Astrophysique, CNRS,
 98 bis Blvd Arago, F-75014 Paris, France
}

\date{Received , Accepted...}

\titlerunning{A Catalogue of Symbiotic Stars}
\authorrunning{Belczy{\'n}ski et al.}

\maketitle

\begin{abstract}

We present a new catalogue of symbiotic stars.  
In our list we include 188 symbiotic stars as well as 28 objects 
suspected of being symbiotic. For each star, we present
basic observational material: coordinates, $V$ and $K$ magnitudes,
ultraviolet (UV), infrared (IR), X-ray and radio observations. We also
list the spectral type of the cool component, the maximum ionization
potential observed, references to finding charts, spectra,
classifications and recent papers discussing the physical parameters
and nature of each object. Moreover, we present the orbital
photometric ephemerides and orbital elements of known symbiotic
binaries, pulsational periods for symbiotic Miras, Hipparcos
parallaxes and information about outbursts and flickering.

\keywords{general: catalogues -- stars: binaries: symbiotic: -- 
          stars: fundamental parameters: classification}

\end{abstract}

\section{Introduction}

Symbiotic stars are interacting binaries, in which an evolved giant
transfers material to much hotter, compact companion. In a typical
configuration, a symbiotic binary comprises a red giant transferring
material to a white dwarf via a stellar wind.

Amongst the evidence for this predominant mass-transfer mechanism
is the fact that ellipsoidal light variations, characteristic of
tidally distorted stars, are rarely observed for symbiotic stars. Thus
far, only two systems, T~CrB (\citetb{46}) and CI~Cyg (\citetb{m229}),
are known to have the ellipsoidal light variations of a distorted
giant.  In some symbiotic systems, the red giant is replaced by a
yellow giant or a carbon star, and the white dwarf by a main-sequence
or neutron star.

Most symbiotic stars ($\sim 80\%$) contain a normal giant star and
these, based on their near-IR colours (showing the presence of stellar
photospheres, $T_{\rm eff} \sim$\ 3000 -- 4000\,{\sc k}), are
classified as S-type systems ({\it stellar}). The remainder contain
Mira variables and their near-IR colours indicate temperatures of
$\sim 1000$\,{\sc k}, giving away the presence of warm dust shells;
these are classified as D-type systems ({\it dusty}). The IR type
seems to be dependent on the orbital separation of the components. For
large separations (long periods), the cool star seems able to evolve
to the Mira stage and produce a dust shell that enshrouds the system; for
smaller separations (shorter periods), we deal with normal giants. For
a detailed review of symbiotic stars, we refer the reader to
\citetb{358}.

Two catalogues of symbiotic stars have been published. The first was
by David Allen in 1984 (\citetb{30}); it included 129 symbiotic stars
and 15 possible symbiotic objects with a concise summary of available
observational data, finding charts and optical spectra for the most of
listed objects. The second catalogue was by Scott Kenyon in 1986
(\citetb{31}); it included 133 symbiotic stars and 20 possible
symbiotic objects, as well as tables describing selected observational
properties of all the objects and a spectroscopic summary of a
selected sample. Kenyon's work also provides the reader with an
excellent overview and bibliography of selected symbiotic stars.

Since 1986, a number of papers have presented surveys of large samples
of symbiotic stars (e.g.\ \citetb{7,vwds93,ibm94,sih95,79,3,313}) and 
in-depth investigations (\citetb{ii88,37,ibesm93,sk94,go96} for AX~Per
alone). New stars have been included in the family of symbiotic stars
each year and, at the same time, better data have been collected and
better data analysis has been performed for a number of well-known
symbiotic stars. The aim of this work is to present the symbiotic star
research community with a comprehensive compilation of existing data
collected from a number of astronomical journals, electronic databases
and unpublished data resources. For many objects a new classification
has been necessary: some have been confirmed as symbiotic stars; some
have been rejected; some new objects have been added. Our catalogue 
lists 188 symbiotic stars and 28 objects suspected of being symbiotic 
stars.

\section{Classification criteria}

The optical spectra of symbiotic stars are characterized by the
presence of absorption features and continuum, as appropriate for a
late-type M giant (often a Mira or semi-regular, SR, variable), and
strong nebular emission lines of Balmer H\,{\sc i}, He\,{\sc ii} and
forbidden lines of [O\,{\sc iii}], [Ne\,{\sc iii}], [Ne\,{\sc v}] and
[Fe\,{\sc vii}]. Some symbiotics -- the yellow symbiotic stars --
contain K (or even G) giants or bright giants. The spectra of many
symbiotic systems also show two broad emission features at $\lambda\,
6825\, {\rm \AA}$ and $\lambda\, 7082\, {\rm \AA}$.  These features
have never been observed in any other astrophysical objects --- only
symbiotic stars with high-excitation nebulae. For many years there was
no plausible identification for these lines, but \citetb{sch89}\
pointed out that the $\lambda\lambda$ 6825, 7082 lines are probably
due to Raman scattering of the O\,{\sc vi} $\lambda\lambda$ 1032, 1038
resonance lines by neutral hydrogen.

To classify an object as symbiotic star we adopted the following
criteria:

\begin{enumerate}
\item The presence of the absorption features of a {\it late-type
giant}; in practice, these include (amongst others) TiO, $\rm H_2O$,
CO, CN and VO bands, as well as Ca\,{\sc i}, Ca\,{\sc ii}, Fe\,{\sc i}
and Na\,{\sc i} absorption lines.
\item The presence of strong emission lines of H\,{\sc i} 
and He\,{\sc i} and either
 \begin{itemize}
  \item emission lines of ions with an ionization potential 
  of at least 35\,eV (e.g.\ [O\,{\sc iii}]), or
  \item an A- or F-type continuum with additional shell absorption lines
  from H\,{\sc i}, He\,{\sc i}, and singly-ionized metals.
 \end{itemize}  
 The latter corresponds to the appearance of a symbiotic star
 in outburst.
\item The presence of the $\lambda$ 6825 emission feature, even
if no features of the cool star (e.g.\ TiO bands) are found.
\end{enumerate} 

Our adopted criteria represent a compromise: a collection of the
classification criteria proposed in the past 70 years (see Kenyon 1986
for details), based on the examples of well-studied and widely
accepted symbiotic objects.  We believe that such an approach is
appropriate, especially given that symbiotic stars are variables with
timescales often exceeding a dozen years and that --- as Kenyon very
sensibly noted --- ``every known symbiotic star has, at one time or
another, violated {\it all} the classification criteria invented''.
For those who would prefer additional or different definitions, we
give the highest ionization potential ever observed in the optical and
UV (for objects that have been observed at least once with the {\em
International Ultraviolet Explorer --- IUE}).  We also comment on all
objects for which our classification may not seem obvious (e.g.\
V934~Her, which some readers may consider to be symbiotic, but which
in our catalogue is classified as a suspected symbiotic star).

\section{The catalogue}

The main catalogue is presented in Table 1.  This table includes
collated data for all the symbiotic stars we know of. Note that a
colon indicates an uncertain measurement or an estimate. Stars are
ordered by right ascension (R.\,A.) for the epoch J2000.0.  The
content of each column is described below.

\noindent
{\bf 1:} Symbiotic star catalogue number. A star symbol, if present
here, means that there is a classification note and/or comment for the
given object. We would still advise the use of the symbiotic (or
suspected symbiotic) star name, as given in the second column of
Tables~1 and 2, and not the object's catalogue number.

\noindent
{\bf 2:} Symbiotic star name. If, for a given object, a variable star
name exists, then it was chosen; otherwise, the name used most often
in the literature was adopted.

\noindent
{\bf 3,4:} R.\,A.\ and declination (J2000.0), taken from radio VLA 
positions (\citetb{m3,9,8}) if available, and if not from the SIMBAD
database but corrected in a few cases were obvious mistakes have been
spotted. If position was taken from somewhere else then comment is 
given in section 4.

\noindent
{\bf 5,6:} Galactic coordinates (not included for extragalactic
objects).

\noindent
{\bf 7,8:} Magnitudes in the $V$ and $K$ filters, respectively.  As
most (if not all) symbiotics are variable, these values are arbitrary
(usually the average of published measurements) just to give the
general level of an object's brightness.

\noindent
{\bf 9:} IR type. If two IR types are given for one object, we supply
references to both estimates in the notes.

\noindent
{\bf 10:} Information on whether an {\em IUE} spectrum (or spectra) is
(are) available {\mbox (+) or not {\mbox ($-$)}.  The number of spectra
can be readily obtained from SIMBAD
(http://simbad.u-strasbg.fr/Simbad) and the spectra can be obtained
from the {\em IUE} data archives (http://nssdca.gsfc.nasa.gov/ndads).
  
\noindent
{\bf 11:} Information about whether an object was ever detected in
X-rays. Plus (+) means a detection; minus ($-$) means that an object
was observed but not detected and that only an upper limit is
available.  Most of the detections and upper limits came from {\em
ROSAT} and were reported in \citetb{275} but some have also been
observed by {\em Einstein} (\citetb{364}), {\em EXOSAT} and {\em
ASCA}.

\noindent
{\bf 12:} Highest ionization potential ever observed in the
emission-line spectra of an object. The potential is given in
electron volts (eV).

\noindent
{\bf 13:} The symbiotic star catalogue number (repeated).

\noindent
{\bf 14:} An estimate of the spectral type of the cool component, with
references.  Since the blue and visual spectral regions are often
contaminated by the circumstellar nebula and/or the hot component, we
have given priority to estimates made in the near-IR region and, in
the case of multiple estimates, to those made at quiescence and/or
near to inferior conjunction of the cool giant.  The estimates based
on the TiO bands are separated from those based on CO 2.3-$\mu$m bands
by `/'.  Also, if the cool component was reported to behave as a Mira
(i.e.\ if Mira-like pulsations have been detected or the object's
position in the near-IR/{\em IRAS} colour diagram coincides with the
region occupied by Mira variables) then it is noted in this column.

\noindent
{\bf 15:} Radio observations of symbiotics. Detections or upper limits
are given. In parentheses, the wavelength of observation is reported.
If more than one detection has been reported, only one is included and
the priority is given to the most extensive radio survey of symbiotics
at 3.6\,cm (\citetb{7}).  Other extensive surveys of symbiotic stars
which were searched for radio detection include \citetb{8,11,9,10,m3}.

\noindent
{\bf 16,17:} {\em IRAS} fluxes at 12 and 25\,$\mu$m.  The fluxes are
taken from pointed observations, if available, (\citetb{13}) or from
survey observations as listed in SIMBAD.  If, for some object, there
was no report of observations in either of the above two sources, but
{\em IRAS} fluxes were available from somewhere else, then the
reference to reported observations is given in the notes. The upper
limits are marked with capital L. The {\em IRAS} number is listed in
Table~8 (which contains different object names for the symbiotic and
suspected symbiotic stars). If the number is not there then the
reference to the reported observations is given in the section
containing comments.

\noindent
{\bf 18:} Major literature references to the object. A number
indicates the reference number; abbreviations in parentheses mark the
subject the reference was noted for: fc -- finding chart, spc --
optical spectrum, class -- classification, parm -- the latest or the
most extensive and up-to-date discussion of an object.

In Table~2, we present data for objects suspected of being symbiotic
stars. The order and content of the columns is exactly the same as in
Table~1.  The catalogue numbers of suspected symbiotic stars are
preceded by the letter `s' throughout the catalogue.

The next two tables include data on symbiotic and suspected symbiotic
star orbits. In Table~3, we have put orbital photometric ephemerides,
including information on the presence of eclipses if available, and
references to every ephemeris estimate. In Table~4, the reader will
find the orbital elements of twenty symbiotic stars as well as
spectroscopic periods, radial velocity semi-amplitudes for the cool 
components, mass ratios, systemic velocities, eccentricities, times 
of inferior spectroscopic conjunctions of the giant, sizes of the 
giant orbits, mass functions and references to each orbital estimate.

In Table~5, we have collected the pulsation ephemerides for Miras in
symbiotic and suspected symbiotic stars.  Again, a reference to every
estimate is given.

Table~6 includes known Hipparcos parallaxes for symbiotic stars.

Table~7 includes information on symbiotic and suspected symbiotic 
star flickering and outburst characteristics.  

Tables~8 includes most of different names for symbiotic and suspected
symbiotic stars.  Symbiotic stars appear first, then suspected
symbiotic stars follow.  Objects are first listed by their catalogue
number, then by the name (translated to SIMBAD nomenclature, if
possible --- the name by which the object is known in Table~1 or 2),
then other names are given.  The names are compatible with SIMBAD and
general internet database nomenclature.  In some cases, the catalogue
name differs between Table~1 (or 2) and Table~8.  This discrepancy is
due the most commonly accepted name (Table~1 or 2) {\em not} following
SIMBAD nomenclature (Table~8).

\section{Classification notes and comments on particular objects}

\begin{center}
Symbiotic stars
\end{center}

\noindent
{\bf 004=SMC3} 
$V$ magnitude during outburst.

\noindent
{\bf 005=SMC N60} 
IR-type S --\citetb{31,35},D --\citetb{30}.

\noindent
{\bf 008=AX Per}
Incorrect coordinates given by \citetb{30,31}.

\noindent
{\bf 009=V471 Per}
This star appears in previous symbiotic catalogs (\citetb{30,31}) as 
V741~Per. The correct name is V471~Per as given in the {\it General Catalog 
of Variable Stars} (\citetb{357}). 

\noindent
{\bf 010=o Ceti}
Cool component is Mira \citetb{22} of type M2-7 III \citetb{201}, UV
spectrum shows emission lines with ionization potentials up to
54.4\,eV \citetb{202} and in the optical spectrum there are emission
lines of H\,{\sc i} and He\,{\sc i} \citetb{204}.  Preliminary orbit
(orbital period = 400 yrs) \citetb{202}.

\noindent
{\bf 011=BD Cam}
Cool component is S giant of type S5.3 \citetb{207}; UV spectra shows
emission lines with ionization potential up to 77.5\,eV \citetb{317}.
24.76-day periodicity estimated from $BVRI$ photometry; pulsational
origin has been suggested \citetb{206}.

\noindent
{\bf 016=UV Aur}
IR-type S --\citetb{31,10},D' --\citetb{30,4}.

\noindent
{\bf 017=V1261 Ori}
Cool component is S giant of type S4.1 \citetb{211};
UV spectrum shows emission lines with ionization potential up to
77.5\,eV \citetb{212}.

\noindent
{\bf 018=LMC1}
{\em IUE} spectra described in \citetb{35}.

\noindent
{\bf 020=Sanduleak's star}
In the optical spectrum, there is an emission feature at 6825 \AA\
\citetb{34}; moreover, there are emission lines with an ionization
potential up to 108.8\,eV \citetb{34,310} including lines of H\,{\sc
i} and He{\sc i} \citetb{35}.  The {\em IUE} spectra are described in
\citetb{35}.

\noindent
{\bf 023=BX Mon}
{\em IRAS} data from \citetb{13}.

\noindent
{\bf 024=V694 Mon}
Object in permanent outburst \citetb{293}; contains M3-5 giant
\citetb{223,m16}; optical spectrum shows emission lines of H\,{\sc i}
and He\,{\sc i} with highly blueshifted ($\sim 2000$--$7000$\,km
s$^{-1}$) shell absorption \citetb{m16,223,m222} and emission lines
of singly-ionised metals \citetb{223} over an A-B type continuum
\citetb{223}.  $VK$ magnitudes are appropriate for the outburst.

\noindent
{\bf 026=RX Pup} 
Highly variable radio emission \citetb{is94}.
Nebula resolved at optical and radio wavelengths with
a possible jet-like feature in the [N\,{\sc ii}] line (\citetb{not3} and
references therein).

\noindent
{\bf 027=Hen 3$-$160}
{\em IRAS} data from \citetb{13}.

\noindent
{\bf 028=AS 201} 
A spherical nebula detected in H$_\alpha$ and [N\,{\sc ii}] lines
(\citetb{not3} and references therein).

\noindent
{\bf 029=KM Vel} 
Cool component is Mira \citetb{22,m49} of M spectral type \citetb{m1};
optical spectrum shows emission lines with ionization potential up to
41.0\,eV \citetb{m121} and emission lines of H\,{\sc i} and He\,{\sc
i} \citetb{121}.  Finding chart in \citetb{185} is incorrect and no
other has been published.

\noindent
{\bf 032=SS73 29} 
{\em IUE} observations reported in \citetb{66}. 
{\em IRAS} data from \citetb{13}.

\noindent
{\bf 033=SY Mus} 
Spectropolarimetric orbit derived in \citetb{not9}.

\noindent
{\bf 034=BI Cru}
A bipolar nebula resolved in the optical (\citetb{not3} and references
therein) with the bipolar lobes and associated outflows perpendicular
to the position angle of intrinsic scattering polarization
\citetb{65}.

\noindent
{\bf 036=TX CVn} 
Low ionization potential (IP$_{\rm max} = 13.6$\,eV), but this is a
confirmed symbiotic star (\citetb{360}: combination spectrum of late B
+ early M, emission lines of H\,{\sc i} and singly-ionised metals).
Classification is also based on its light curve showing eruptions as
in other symbiotics (with $\Delta m_{pg}$ up to $\sim 3^m$).  Since
the 1970's, the star is in permanent outburst with P-Cyg type
spectrum.

\noindent
{\bf 038=Hen 3$-$828} 
{\em IRAS} data from \citetb{13}.

\noindent
{\bf 041=St 2$-$22} 
The SIMBAD database uses different name for this object: PN Sa 3$-$22.

\noindent
{\bf 043=V840 Cen} 
{\em IRAS} data from \citetb{118}. 
Finding chart available in \citetb{237} where object is marked as star A 
\citetb{186}.

\noindent
{\bf 046=Hen 3$-$916} 
Finding chart in \citetb{30} is wrong, object is 2mm ($\sim 20"$) 
E of marked star \citetb{52}.

\noindent
{\bf 047=V704 Cen} 
Cool component might be Mira \citetb{22}.

\noindent
{\bf 048=V852 Cen}
Cool component is Mira \citetb{22,m49}; optical spectrum shows
emission lines with ionization potential up to 100\,eV
\citetb{57,175,3}; moreover, optical spectrum shows emission
feature at 6825 \AA\ \citetb{m126} and emission lines of H\,{\sc i}
and He\,{\sc i} \citetb{3}. Bipolar nebula resolved in the optical
(Southern Crab) (\citetb{not3} and references therein).

\noindent
{\bf 050=V417 Cen}
An irregular nebula resolved at optical wavelengths (\citetb{not3} and
references therein).

\noindent
{\bf 055=HD 330036} 
This is a yellow symbiotic star; cool component is F5 giant or subgiant 
(\citetb{62}). 
In UV, there are emission lines with ionization potential up to 
77.5\,eV \citetb{62} and in the optical spectrum there are emission lines with 
ionization potential up to 54.4\,eV \citetb{3,62}. 
IR-type D' --\citetb{0}.

\noindent
{\bf 056=Hen 2$-$139} 
Only H\,{\sc i} emission lines in spectrum according to \citetb{30}, but 
other emission lines (like [O\,{\sc iii}]) are reported in \citetb{m121}.

\noindent
{\bf 058=AG Dra} 
A secondary periodicity of $\sim 355^{\rm d}$\ has been detected
in the optical light curve and interpreted in terms of non-radial
pulsation of the cool giant \citetb{not1}.
An orbital inclination, $i \sim 120^{\circ}$, has been derived from 
spectropolarimetric observations \citetb{not2}.

\noindent
{\bf 060=V347 Nor}
An elliptical nebula resolved at optical wavelengths (\citetb{not3} and
references therein).

\noindent
{\bf 065=Hen 3$-$1213} 
{\em IUE} observations reported in \citetb{90}.

\noindent
{\bf 066=Hen 2$-$173} 
{\em IRAS} data from \citetb{13}.

\noindent
{\bf 067=Hen 2$-$176} 
IR-type S --\citetb{3},D --\citetb{30,31}.

\noindent
{\bf 068=KX TrA} 
The finding chart in \citetb{30} is wrong: the object is really 3mm 
($\sim 25"$) W of marked star, although tabulated coordinates are correct 
\citetb{52}.

\noindent
{\bf 071=CL Sco} 
{\em IRAS} data from \citetb{78}.

\noindent
{\bf 073=V455 Sco}
An elliptical nebula possibly resolved in [O\,{\sc iii}] (\citetb{not3} and
references therein).

\noindent
{\bf 074=Hen 3$-$1341} 
{\em IUE} observations reported in \citetb{90}.
Spectral signatures of collimated bipolar jets have been found during the 
1999 outburst \citetb{m230}.

\noindent
{\bf 077=H 2$-$5} 
IR-type D --\citetb{3},S --\citetb{30,31}.

\noindent
{\bf 084=V2116 Oph} 
Orbital period of 303.8 days is derived from the spin changes 
of the X-ray pulsar companion \citetb{pbj99,cdd86}.

\noindent
{\bf 088=M 1$-$21} 
$VK$ magnitudes -- close and fainter companion also measured.

\noindent
{\bf 089=Hen$-$251} 
$K$-band spectrum is practically identical with that of the 
symbiotic Mira, RX Pup, as observed during the dust obscuration event, with 
strong dust continuum and weak CO 2.3-$\mu$m band \citetb{ntt00,mik00}.

\noindent
{\bf 092=RT Ser} 
{\em IRAS} data from \citetb{13}.

\noindent
{\bf 093=AE Ara} 
{\em IRAS} data from \citetb{13}.

\noindent
{\bf 094=SS73 96} 
{\em IRAS} data from \citetb{13}.
An axisymmetrical nebula resolved at radio wavelengths (\citetb{not3} and
references therein).

\noindent
{\bf 096=V2110 Oph} 
{\em IRAS} data from \citetb{13}.

\noindent
{\bf 100=H 1$-$36}
In \citetb{5} there is an estimate of cool component spectral type
M4-5 III based on TiO 7100\AA\ band depth. However the spectrum of H
1$-$36 shown on their Fig.A1 does not show any absorption features or
red continuum. A complex nebula resolved at optical and radio
wavelengths (\citetb{not3} and references therein). The only symbiotic
star known to support an OH maser (\citetb{ish94}).

\noindent
{\bf 101=RS Oph} 
Bipolar nebula detected in radio range (\citetb{not3} and references
therein).

\noindent
{\bf 102=WRAY 16$-$312} 
{\em IRAS} and $JHKL$ colours confirm earlier suggestions \citetb{m2,22}
that cool component of this system is a Mira \citetb{0}. 
In the optical spectrum presented in \citetb{30} there are emission lines with 
ionization potential up to 108.8\,eV and moreover lines of H\,{\sc i} and He\,{\sc i} are 
present \citetb{0}.
{\em IRAS} data from \citetb{13}.

\noindent
{\bf 103=V4141 Sgr} 
Classified as S-type in \citetb{30,4}, but in the near-IR/{\em IRAS} colour 
diagrams it falls in the region occupied by symbiotic Miras \citetb{3}. 
$K$-band spectrum shows strong CO 2.3-$\mu$m band consistent with an M6 giant 
\citetb{ntt00}. 
Spectral type of cool component also estimated in \citetb{179,30} 
to be mid or late M.

\noindent
{\bf 105=AS 245} 
Classified as S-type in \citetb{16,3} but in the near-IR/{\em IRAS} colour 
diagrams it falls in the region occupied by symbiotic Miras \citetb{0}.

\noindent
{\bf 107=Bl 3$-$14}
The finding chart in \citetb{30} is good, but the coordinates are reported to 
disagree with the measured position: $\alpha=17^{\rm h} 52^{\rm m} 06^{\rm s}.4, 
\delta= -29^{\rm \circ} 45' 49''$ 
(1950) \citetb{52} (if this is right, our coordinates should also be 
corrected).

\noindent
{\bf 110=V745 Sco} 
$VK$ magnitudes during decline from outburst \citetb{253}.

\noindent
{\bf 112=AS 255} 
IR-type S --\citetb{30,44},D --\citetb{31,43}.

\noindent
{\bf 114=H2$-$34} 
Spectral type M5 is estimated by comparing  `by eye' the depths of TiO 
$\lambda$ 6180 and $\lambda$ 7100 $\rm \AA$ in the spectrum in Fig.~2 
in \citetb{3} with those of spectral standards.

\noindent
{\bf 115=SS73 117} 
{\em IRAS} data from \citetb{13}.

\noindent
{\bf 116=AS 269}
This is a yellow symbiotic star, cool component is G-K giant \citetb{3,m22}. 
In the optical spectrum there are emission lines with ionization
potential up to 54.4\,eV \citetb{184}.

\noindent
{\bf 118=SS73 122} 
IR-type D --\citetb{13}, others note only possible S type (\citetb{30,31}).     

\noindent
{\bf 120=H 2$-$38}
There was a report of a pulsational period of 433 days for this star
in \citetb{117}, but this is a mistake and the reported number is the
pulsation period of another symbiotic star: V366 Car (Hen 2$-$38).
The spectral type of the cool component is estimated in \citetb{16} to
be M8.5.

\noindent
{\bf 122=Hen 3$-$1591} 
IR-type D --\citetb{30,4}, S --\citetb{31,m73}.

\noindent
{\bf 124=Ve 2$-$57} 
Cool component is M star \citetb{m16}. 
In the optical spectrum there are emission lines with ionization potential up 
to 35.1\,eV or probably up to 54.4\,eV \citetb{m16}.

\noindent
{\bf 125=AS 276} 
IR-type S --\citetb{30,44},D --\citetb{31}. 
There is also a D' classification in \citetb{43}, but it doesn't look reliable.

\noindent
{\bf 128=V2506 Sgr}
{\em IRAS} data from \citetb{108}.

\noindent
{\bf 132=YY Her} 
{\em IRAS} data from \citetb{13}.

\noindent
{\bf 133=V2756 Sgr}
Finding chart in \citetb{185} is incorrect (\citetb{m23}).  

\noindent
{\bf 134=FG Ser} 
$K$ magnitude during outburst. 
Coordinates taken from \citetb{m233} -- SIMBAD coordinates are not correct.

\noindent
{\bf 138=V4074 Sgr} 
{\em IUE} observations reported in \citetb{79}.

\noindent
{\bf 139=V2905 Sgr} 
{\em IRAS} data from \citetb{78}. 
Spectral type of cool component also estimated in \citetb{291} to be K/M.

\noindent
{\bf 146=V3811 Sgr} 
Mis-identified in \citetb{185} and in \citetb{m42} (see \citetb{m73}).

\noindent
{\bf 148=V3890 Sgr}
Cool component is M4-8 giant (\citetb{m96,261,262}). 
In the optical spectrum there are emission lines with an ionization potential 
up to 361\,eV \citetb{261}.
This object was earlier classified as recurrent nova with M type companion
\citetb{262,m96}.
The spectrum is also presented in \citetb{254}.

\noindent
{\bf 156=FN Sgr} 
{\em IRAS} data from \citetb{78}.

\noindent
{\bf 160=V1413 Aql}
Spectral type M4 estimated from the TiO $\lambda$ 7100 band depth as 
observed during mid-eclipse \citetb{mik00a}.

\noindent
{\bf 162=Ap 3$-$1}
Short description of optical spectrum is given in \citetb{30}.
The object was identified with the 2U 1907+2 X ray source \citetb{m104} but so far 
there is no {\em ROSAT} detection, so this identification might not be correct.

\noindent
{\bf 166=BF Cyg} 
{\em IRAS} data from \citetb{13}.

\noindent
{\bf 167=CH Cyg}
Complex nebula with jet-like features resolved at optical and radio
wavelengths (\citetb{not3} and references therein).  Both the light
curves and the radial velocity curves show multiple periodicities: a
$\sim 100^{d}$\ photometric period has been attributed to radial
pulsation of the giant \citetb{m107}, while the nature of the
secondary period of $\sim 756^{\rm d}$\, also present in the radial
velocity curve, is not clear \citetb{not5}.  There is controversy
about whether the system is triple or binary \citetb{93}, and whether
the symbiotic pair is the inner binary \citetb{not6} or the white
dwarf is on the longer orbit \citetb{not7,not8}.

\noindent
{\bf 169=HM Sge} 
Mean $K$ magnitude during outburst.
A complex nebula with possible jet-like features resolved
at optical and radio wavelengths (\citetb{not3} and references therein).
The nebula is aligned with the binary orientation deduced from
spectropolarimetry of the Raman scattered O\,{\sc vi} lines \citetb{m240}.

\noindent
{\bf 170=Hen 3$-$1761} 
{\em IRAS} data from \citetb{13}. 
{\em IUE} observations reported in \citetb{90}.

\noindent
{\bf 171=QW Sge}
{\em IRAS} data from \citetb{13} although \citetb{145} report no {\em IRAS} detection.

\noindent
{\bf 172=CI Cyg}
Coordinates from VLA observations \citetb{m232}.

\noindent
{\bf 174=V1016 Cyg} 
A complex nebula with possible jet-like features resolved
at optical and radio wavelengths (\citetb{bang,not3} and references therein).

\noindent
{\bf 176=PU Vul} 
$V$ mag during the decline from outburst (XI 1994) \citetb{153}.
In \citetb{m231} $\sim 211^d$ periodicity has been reported.

\noindent
{\bf 177=LT Del}
{\em IRAS} data from \citetb{78}.
Spectral type of cool component also estimated in \citetb{m24} to be G5.

\noindent
{\bf 178=V1329 Cyg}
Spectral type of cool component also estimated in \citetb{36} to be $>$M4.

\noindent
{\bf 180=ER Del}
Cool component is S star of type S5.5/2.5 \citetb{267}.  In the UV,
there are emission lines with an ionization potential up to 47.9\,eV
and a strong UV continuum indicates the presence of a hot companion
\citetb{268}; in the optical spectrum there are emission lines of
H\,{\sc i} \citetb{268}.

\noindent
{\bf 181=V1329 Cyg}
The system inclination, $i=86^{\circ} \pm 2^{\circ}$, and
the position angle of the orbital plane, $11^{\circ}$, has been
derived from spectropolarimetric studies. 
An extended nebulosity detected in the [O\,{\sc iii}] $\lambda 5007$\ line is aligned  
with the orbital plane \citetb{159}.

\noindent
{\bf 183=V407 Cyg} 
{\em IRAS} data from \citetb{118}. 
IR-type S --\citetb{m3}, D --\citetb{18} and also there is D' estimate in 
\citetb{10}.
In \citetb{18} there is an estimate of the orbital period of 43 yrs.

\noindent
{\bf 184=StHA 190} 
In \citetb{59} there is a suggestion, based on the {\em IRAS} ratio of
F$_{12}$/F$_{25}$, that the cool component in this system is a Mira
variable.  Comparison of {\em IRAS} fluxes with diagnostic diagrams in
\citetb{13} shows that this object is among or close to D' systems,
and the $VJHKL$ colours are consistent with a G-K giant, so there is
no reason to think that a Mira variable is present in this binary.
The authors of \citetb{59} argue that F$_{12}$/F$_{25}>1.0$ suggests
the presence of a Mira but it may be merely the signature of dust
around the system.

\noindent
{\bf 185=AG Peg} 
$VK$ magnitudes during outburst. 
A complex nebula with possible bipolar structure
detected at optical and radio wavelengths (\citetb{not3} and references
therein).

\noindent
{\bf 186=LL Cas}
The presence of the [Fe\,{\sc vii}] 4892\AA\ line reported in
\citetb{274} is not reliable because of the absence of the strongest
[Fe\,{\sc vii}] 6087\AA\ iron line at that time.  In \citetb{274},
there is a report of a possible pulsational period for the cool
component of this system (286.6 days). This is a plausible
explanation, as the spectrum taken at maximum light shows a more
pronounced late-type continuum than the spectrum taken at minimum (see
\citetb{274}), indicating that the cool component is responsible for
this variability. IR colours: J$=8.90,H=8.03,K=7.55$ \citetb{m103})
with assumed modest amount of interstellar reddening (A$_{\rm K}$=0.2)
give $J_0=8.44,H_0=7.67,K_0=7.35$ which corresponds to the colours of
a normal giant in an S-type symbiotic star, although this might still
be a Mira without an IR excess (like the Mira in R~Aqr, which is
another S-type symbiotic star).

\noindent
{\bf 187=Z And}
Spectral type of cool component also estimated in \citetb{m90} to be 
$\sim$M5.2.
An inclination of $i=47 \pm 12^{\circ}$\ and an orbit orientation,
$\Omega=72 \pm 6^{\circ}$, derived from spectropolarimetry \citetb{not12}.

\noindent
{\bf 188=R Aqr}
The binary has been spatially resolved and a preliminary orbit (with a
period of $\sim 44$ yrs) derived in \citetb{hpl97}.  The system is
embedded in a complex bipolar nebula with jets (\citetb{not3} and
references therein). Only symbiotic star known to support H$_2$O and
SiO masers (\citetb{ish94,m1}).

\begin{center}
Suspected symbiotic stars
\end{center}

\noindent
{\bf s01=RAW 1691}
Carbon star \citetb{199} + H$_\alpha$\ profile as for interacting binary 
star \citetb{m18}.

\noindent
{\bf s02=[BE74] 583} 
Suspected in \citetb{m139}.

\noindent
{\bf s03=StHA 55}
Carbon star \citetb{61} + with strong H\,{\sc i} emission \citetb{61} 
(too strong for single carbon star).

\noindent
{\bf s04=GH Gem}
Suspected in \citetb{178,31}.

\noindent
{\bf s05=ZZ CMi} 
This object was classified as symbiotic in \citetb{48,222}.  We
disagree with this classification because: {\it i)} colours are bluer
at minimum \citetb{316}, the opposite to what is observed for
symbiotics; the light curve looks more like a pulsational curve and
not like a symbiotic light curve; {\it ii)} the spectrum presented in
\citetb{48} does not look like a symbiotic spectrum (e.g.\ H$_\gamma >
{\rm H}_\beta$) and is noisy ([Ne\,{\sc iii}] line may not be present
(so IP$_{\rm max}$=35.1\,eV).  However, this object contains a
late-type star (though we do not know if the star is giant) and it
displays an emission-line spectrum; also, the H$_\alpha$ profile shown
in \citetb{222}) looks like a symbiotic star (for comparison see
\citetb{m18}). We therefore include this object as suspected
symbiotic.

\noindent
{\bf s06=NQ Gem}
Suspected in \citetb{225}. Highly variable UV continuum with strong 
C\,{\sc iv}] emission and Si\,{\sc iii}]/C\,{\sc iii}] ratio similar to symbiotic stars. 
He\,{\sc ii} 1640\AA\ emission line has been detected in 1979 by {\em IUE}.

\noindent
{\bf s07=WRAY 16$-$51} 
Probable presence of late-type star and emission-type spectrum (H\,{\sc i} 
emission lines) \citetb{179}.

\noindent
{\bf s08=Hen 3$-$653}
Suspected in \citetb{30,291}: late-type star and
emission-type spectrum (H\,{\sc i} and He\,{\sc i} emission lines).

\noindent
{\bf s09=NSV 05572} 
Late-type giant and emission type-spectrum (H\,{\sc i} emission lines).

\noindent
{\bf s10=AE Cir} 
Suspected in \citetb{240}.
Periods of 3900 and 100 days are mentioned in \citetb{240} 
(based on visual photometric observations).

\noindent
{\bf s11=V748 Cen} 
Suspected in \citetb{31}: M type giant \citetb{177,55} and
emission-line spectrum (H\,{\sc i}, Fe\,{\sc ii}, [Fe\,{\sc
ii}], [S\,{\sc ii}]) \citetb{300} and UV excess.

\noindent
{\bf s12=V345 Nor}
Suspected in \citetb{m149}: M star \citetb{245} and 
emission-line spectrum (H\,{\sc i}, Fe\,{\sc ii}) \citetb{245}.
Listed as N Nor 1985/2 in \citetb{131}.

\noindent
{\bf s13=V934 Her}
Suspected in \citetb{250}: M bright giant and UV emission lines with
ionization potential up to 77.5\,eV but no emission lines in optical
spectrum and no short-wavelength continuum was found (the
1200--2800\AA\ integrated flux $< 1.5 \times 10^{-14}$ erg s$^{-1}$
cm$^{-2}$ \AA$^{-1}$ at Earth) which excludes the presence of a hot
white dwarf companion (although a neutron star is still possible).

\noindent
{\bf s14=Hen 3$-$1383} 
Possible M type star \citetb{m19} and emission-type spectrum 
(H\,{\sc i}, He\,{\sc i}) \citetb{m16}.
Nebula resolved at radio wavelengths (\citetb{not3} and references therein).

\noindent
{\bf s15=V503 Her}
Suspected in \citetb{31}: M star \citetb{m119} and blue
excess in the optical spectra suggesting presence of hot companion.

\noindent
{\bf s16=WSTB 19W032} 
Late type giant \citetb{232} and emission-line spectrum:  
lines of H\,{\sc i}, He\,{\sc i} and others with ionization potential up to 35.1\,eV.
But this emission-line spectrum might not be physically connected with
the giant \citetb{232}.

\noindent
{\bf s17=WRAY 16$-$294}
Suspected in \citetb{3}: red continuum typical of reddened K giant 
and emission-line spectrum (H\,{\sc i}, He\,{\sc i} and others with ionization potential 
up to 35.1\,eV).
WRAY 16$-$294 appears as WRAY 16$-$296 in \citetb{3}.

\noindent
{\bf s18=AS 241} 
Suspected in \citetb{30}: M star \citetb{3} and emission-line
spectrum (H\,{\sc i}, He\,{\sc i}) \citetb{3}.
M6 spectral type of cool component and D' IR type from \citetb{43} 
are not reliable, as M6 does not agree with IR colours ($JHK$)
and authors do not follow original definition of D'.

\noindent
{\bf s19=DT Ser}
Considered as symbiotic in \citetb{280,m122}: emission spectrum of
H\,{\sc i}, He\,{\sc i} and other lines with ionization potential up
to 54.4\,eV plus G? \citetb{280} or G2-K0 III-I \citetb{m122} cool
component.  But there is a report of a G star 5$''$ from this object,
so the cool component may not be connected physically with the source
of the emission-line spectrum.

\noindent
{\bf s20=V618 Sgr} 
Presence of late-type component (TiO bands in optical spectrum 
\citetb{279}) and emission-line spectrum (H\,{\sc i}, Fe\,{\sc ii} \citetb{279}).

\noindent
{\bf s21=AS 280} 
Suspected in \citetb{3}: this object resembles a symbiotic star
in outburst.

\noindent
{\bf s22=AS 288} 
This object shows optical emission-line spectrum (H\,{\sc i}, He\,{\sc i} 
and others with ionization potential up to 54.4\,eV) but no late-type
component has been seen so far, however $K$ magnitude and {\em IRAS} fluxes 
compared to diagnostic diagrams in \citetb{13} place this object among 
symbiotics (of IR type D), emission-line fluxes ([O,{\sc iii}]4363, 5007\AA, 
H$_\beta$, H$_\gamma$) compared to diagnostic diagrams in \citetb{m5} 
place this object also among symbiotics (among IR type S but close to 
D-type objects).

\noindent
{\bf s23=Hen 2$-$379}
Cool component is G-K giant \citetb{258} and there is emission-line
spectrum: H\,{\sc i}, He\,{\sc i} and other lines with ionization potential up to 
35.1\,eV.
But K giant might not be physically associated with nebula, which is 
source of the emission \citetb{257}. 
Finding chart in \citetb{185} is unclear as reported in \citetb{m23}.

\noindent
{\bf s24=V335 Vul} 
Suspected in \citetb{287}: presence of carbon giant and optical 
emission-line spectrum (H\,{\sc i}) displaying hot continuum in blue.
We agree with this classification although this object might be a single
pulsating star:
{\it (i)} the carbon star might pulsate with period of 342 days
\citetb{m172} and then emission lines behave as for a Mira variable -- 
they disappear near minimum light and show up again at maximum 
(see spectra in \citetb{287});
{\it (ii)} H$_\alpha$\ is very narrow: 2\AA\ (90\,km\,s$^{-1}$) and for a symbiotic star 
we would expect a width of about 300--500\,km\,s$^{-1}$;
{\it (iii)} the Balmer decrement is different than that observed for symbiotic stars
(it resembles that of Mira variable), although the authors of
\citetb{287} claim that the Balmer decrement resembles that of a symbiotic
star.

\noindent
{\bf s25=V850 Aql} 
Probable presence of Mira \citetb{m41,357} or late-type star
\citetb{m22} in the centre of planetary nebula PK 037$-$6 2 (see note 
in \citetb{357}) with emission-line spectrum (H\,{\sc i} lines).
In \citetb{m22} and \citetb{m41} there are notes that in \citetb{m42} 
this object is classified as symbiotic, but this is not true and in
\citetb{m42} there are only IR colours for V850 Aql.

\noindent
{\bf s26=Hen 2$-$442} 
Suspected in \citetb{321}: TiO bands probably present, suggesting 
cool component \citetb{m176,321} and optical emission-line spectrum: 
H\,{\sc i}, He\,{\sc i} and other lines with ionization potential up to 100\,eV.
Hen 2$-$442 consists of two PN like objects: Hen 2$-$442A and Hen 2$-$442B 
\citetb{m176} and values in catalogue are for the whole system. 
Symbiotic nature was suggested for Hen 2$-$442A.

\noindent
{\bf s27={\em IRAS} 19558+3333}
Suspected in \citetb{si94}: OH/IR star, based on {\em IRAS} colours, but
without an OH maser, so a probable, extreme D-type system. Radio continuum
emission implies a hot, ionising companion. Correct coordinates given here
for precise radio emission (incorrect coordinates given by \citetb{si94}).

\noindent
{\bf s28=V627 Cas} 
Suspected in \citetb{m192}.
Spectral type of cool component also estimated in \citetb{m190}
to be M2-4.

\section{Comments on other objects not included in the catalogue}

\noindent
{\bf V1017 Sgr} 
In some publications, regarded as symbiotic, probably after 
inclusion in Kenyon's catalogue (\citetb{31}), but this is not a
symbiotic star.
This object is a cataclysmic variable with orbital period of 5.7 days 
(\citetb{129}).

\noindent
{\bf CI Cam}
The optical counterpart of XTE J0421+560.
Reported as symbiotic in \citetb{357}, possibly after suggestion in 
\citetb{m237}.
It is, however, a high-mass X-ray binary with a Be star donor 
(\citetb{m238,m239}).

\begin{acknowledgements}
This work has been funded by the KBN grant 2P03D02112, and also made
use of the NASA Astrophysics Data System and SIMBAD database. 
We would like to thank Dr Estella Brandi and Dr Maciej Miko{\l}ajewski 
for many helpful comments.
RJI acknowledges the award of a PPARC Advanced Fellowship. KB would like
also to thank Dr Tomasz Bulik for help with preparation of this
manuscript.
\end{acknowledgements}

\onecolumn
\newpage

\clearpage

\noindent
{\bf Table 1.} Symbiotic star\vspace*{0.2cm}s\\
\begin{tabular}{|l|c|c|c|c|c|c|c|c|c|c|c|}
\hline \hline
No.& Name& $\alpha$(2000)& $\delta$(2000)& l$^{\rm II}$& b$^{\rm II}$& $V$& $K$& IR& {\em IUE}& X& IP$_{\rm max}$\cr
&&\ $^h$\ \ $^m$\ \ $^s$& \ $^\circ$\ \ '\ \ \ '' &\ $^\circ$&\ $^\circ$& [mag]& [mag]&&&& [eV] \cr
\hline \hline

001&  SMC1& 00 29 10.9&  $-$74 57 38.9& & &  16.2& &  S:&  +&  & 114 \cr 
\hline

002&  SMC2& 00 42 48.1&  $-$74 42 00.0& & &  16.2& &  S:&  +&  $-$& 114 \cr 
\hline

003&  EG And& 00 44 37.1&  +40 40 45.7& 121.54& $-$22.17&  7.1&  2.6&  S&  +&  +&  100 \cr 
\hline

004$^\star$&  SMC3& 00 48 19.9&  $-$73 31 54.9& & &  15.5:& &  S:&  +&  +&  235 \cr 
\hline

005$^\star$&  SMC N60& 00 57 12.0&  $-$74 13 00.0& & &  16.8&  13.0&  S,D&  +&  $-$& 114 \cr 
\hline

006&  LIN 358& 00 59 24.0&  $-$75 04 59.9& & &  15.2&  11.4&  S&  +&  +& 114 \cr
\hline

\end{tabular}

\newpage
\noindent
{\bf Table 1.} Symbiotic stars -- cont\vspace*{0.2cm}.\\
\begin{tabular}{|l|c|c|c|c|c|c|}
\hline \hline
No.& Cool-star spectrum& Radio& {\em IRAS}& {\em IRAS}& References\cr
 & & [mJy]& F$_{12}$[Jy]& F$_{25}$[Jy]& \cr
\hline \hline

001&  C3.2:,C(\cite{35,34})& & & &  \cite{34}(fc,spc,class) \cite{35}(parm)  \cr 
\hline

002&  K,G-K(\cite{35,34})& & & &  \cite{34}(fc,spc,class) \cite{35}(parm)  \cr 
\hline

003&  M3(\cite{ms99})&  0.54(3.6cm)&  4.5&  1.25&  \cite{30}(fc,class) \cite{31}(spc) \cite{26}(parm)  \cr 
\hline

004&  M0,K-M(\cite{34,35})& & & &  \cite{34}(fc,spc,class) \cite{197}(parm)  \cr 
\hline

005&  C3.3(\cite{35})& & & &  \cite{30}(fc,class) \cite{35}(spc,parm)  \cr 
\hline

006&  mid K(\cite{35})& & & &  \cite{30}(fc,class) \cite{35}(spc,parm)  \cr
\hline

\end{tabular}

\newpage
\noindent
{\bf Table 1.} Symbiotic stars -- cont\vspace*{0.2cm}.\\
\begin{tabular}{|l|c|c|c|c|c|c|c|c|c|c|c|}
\hline \hline
No.& Name& $\alpha$(2000)& $\delta$(2000)& l$^{\rm II}$& b$^{\rm II}$& $V$& $K$& IR& {\em IUE}& X& IP$_{\rm max}$\cr
&&\ $^h$\ \ $^m$\ \ $^s$& \ $^\circ$\ \ '\ \ \ '' &\ $^\circ$&\ $^\circ$& [mag]& [mag]&&&& [eV] \cr
\hline \hline

051&  BD$-$21 3873& 14 16 34.3&  $-$21 45 50.2& 327.88& +36.95&  10.7&  7.2&  S&  +&  $-$&  77.5 \cr 
\hline

052&  Hen 2$-$127& 15 24 49.8&  $-$51 49 52.6& 325.53& +4.19&  16.0&  8.1&  D&  &  $-$& 114 \cr 
\hline

053&  Hen 3$-$1092& 15 47 10.6&  $-$66 29 16.0& 319.22& $-$9.35&  13.5&  7.8&  S&  +&  $-$& 114 \cr 
\hline

054&  Hen 3$-$1103& 15 48 28.5&  $-$44 19 00.9& 333.21& +7.90&  13.0&  8.4&  S&  &  $-$& 114 \cr 
\hline

055$^\star$&  HD 330036& 15 51 15.9&  $-$48 44 58.5& 330.78& +4.15&  11.0&  7.6&  D'&  +&  $-$&  77.5 \cr 
\hline

056$^\star$&  Hen 2$-$139& 15 54 45.4&  $-$55 29 36.9& 326.91& $-$1.40&  16.8&  5.9&  D&  &  $-$&  35.1 \cr
\hline

\end{tabular}

\newpage
\noindent
{\bf Table 1.} Symbiotic stars -- cont\vspace*{0.2cm}.\\
\begin{tabular}{|l|c|c|c|c|c|c|}
\hline \hline
No.& Cool-star spectrum& Radio& {\em IRAS}& {\em IRAS}& References\cr
 & & [mJy]& F$_{12}$[Jy]& F$_{25}$[Jy]& \cr
\hline \hline

051&  K2(\cite{ms99})&  $<$0.22(3.6cm)& & &  \cite{30}(fc,class,spc) \cite{81}(parm)  \cr 
\hline

052&  Mira M5/M7(\cite{ms99,16})&  0.70(3.5cm)&  0.49&  0.29&  \cite{30}(fc,class,spc) \cite{3}(parm)  \cr 
\hline

053&  M5.5(\cite{ms99})&  10(6cm)& & &  \cite{30}(fc,class,spc) \cite{3}(parm)  \cr 
\hline

054&  M3.5/M0(\cite{ms99,16})&  $<$1(2cm)& & &  \cite{30}(fc,class,spc)  \cr 
\hline

055&  G5(\cite{ms99})&  34(2cm)&  19.53&  38.58&  \cite{30}(fc,spc) \cite{0}(class) \cite{59}(parm)  \cr 
\hline

056&  Mira,M6.5(\cite{ms99})& &  5.87&  2.99&  \cite{30}(fc,class,spc)  \cr
\hline

\end{tabular}

\newpage
\noindent
{\bf Table 1.} Symbiotic stars -- cont\vspace*{0.2cm}.\\
\begin{tabular}{|l|c|c|c|c|c|c|c|c|c|c|c|}
\hline \hline
No.& Name& $\alpha$(2000)& $\delta$(2000)& l$^{\rm II}$& b$^{\rm II}$& $V$& $K$& IR& {\em IUE}& X& IP$_{\rm max}$\cr
&&\ $^h$\ \ $^m$\ \ $^s$& \ $^\circ$\ \ '\ \ \ '' &\ $^\circ$&\ $^\circ$& [mag]& [mag]&&&& [eV] \cr
\hline \hline

101$^\star$&  RS Oph& 17 50 13.2&  $-$06 42 28.4& 19.80& +10.37&  11.5&  6.5&  S&  +&  +& 114 \cr 
\hline

102$^\star$&  WRAY 16$-$312& 17 50 16.5&  $-$30 57 41.1& 358.79& $-$1.91& &  7.8&  D&  &  +&  108.8 \cr 
\hline

103$^\star$&  V4141 Sgr& 17 50 24.0&  $-$19 53 42.0& 8.31& +3.73&  16.4&  7.9&  D,S&  &  $-$& 54.4 \cr 
\hline

104&  ALS 2& 17 50 51.1&  $-$17 47 56.0& 10.18& +4.70&  14.2& &  S&  &  & 114 \cr 
\hline

105$^\star$&  AS 245& 17 51 00.8&  $-$22 19 41.3& 6.29& +2.38&  11.0&  7.2&  S,D&  &  $-$& 114 \cr 
\hline

106&  Hen 2$-$294& 17 51 38.2&  $-$32 50 57.0& 357.31& $-$3.12& &  8.4&  S&  &  $-$& 114 \cr
\hline

\end{tabular}

\newpage
\noindent
{\bf Table 1.} Symbiotic stars -- cont\vspace*{0.2cm}.\\
\begin{tabular}{|l|c|c|c|c|c|c|}
\hline \hline
No.& Cool-star spectrum& Radio& {\em IRAS}& {\em IRAS}& References\cr
 & & [mJy]& F$_{12}$[Jy]& F$_{25}$[Jy]& \cr
\hline \hline

101&  M0-M2(\cite{105,am99})&  $<$0.33(3.6cm)&  0.30&  0.16&  \cite{30}(fc,class,spc) \cite{105}(parm)  \cr 
\hline

102&  Mira(\cite{22})&  0.96(6cm)&  5.0&  3.1&  \cite{30}(fc,spc) \cite{0}(class)  \cr 
\hline

103&  M6(\cite{ntt00})&  $<$0.17(6cm)&  1.88&  0.65&  \cite{30}(fc) \cite{123}(spc,class) \cite{3}(parm) \cr 
\hline

104&  M2(\cite{3})& & & &  \cite{1}(fc,spc,class) \cite{3}(parm)  \cr 
\hline

105&  Mira(\cite{22});M6(\cite{16})&  0.87(6cm)&  1.67&  0.62&  \cite{30}(fc,class,spc) \cite{3}(parm)  \cr 
\hline

106&  M3(\cite{3})&  $<$0.11(3.6cm)&  2.19&  2.89&  \cite{30}(fc,class,spc) \cite{3}(parm)  \cr
\hline

\end{tabular}

\newpage
\noindent
{\bf Table 1.} Symbiotic stars -- cont\vspace*{0.2cm}.\\
\begin{tabular}{|l|c|c|c|c|c|c|c|c|c|c|c|}
\hline \hline
No.& Name& $\alpha$(2000)& $\delta$(2000)& l$^{\rm II}$& b$^{\rm II}$& $V$& $K$& IR& {\em IUE}& X& IP$_{\rm max}$\cr
&&\ $^h$\ \ $^m$\ \ $^s$& \ $^\circ$\ \ '\ \ \ '' &\ $^\circ$&\ $^\circ$& [mag]& [mag]&&&& [eV] \cr
\hline \hline

151&  AS 316& 18 42 33.0&  $-$21 17 48.0& 12.89& $-$7.67&  12.0&  7.8&  S&  &  $-$& 114 \cr 
\hline

152&  DQ Ser& 18 44 40.0&  +05 03 29.9& 36.67& +3.81&  14.6& & &  &  &  54.4 \cr 
\hline

153&  MWC 960& 18 47 55.9&  $-$20 05 49.9& 14.53& $-$8.27 &  12.0&  7.8&  S&  +&  $-$& 114 \cr 
\hline

154&  AS 323& 18 48 35.7&  $-$06 41 08.9& 26.66& $-$2.41  &  15.5&  7.9&  S&  &  $-$& 114 \cr 
\hline

155&  AS 327& 18 53 16.9&  $-$24 22 54.0& 11.14& $-$11.23&  11.8&  8.5&  S&  &  $-$& 114 \cr 
\hline

156$^\star$&  FN Sgr& 18 53 52.9&  $-$18 59 42.0& 16.15& $-$9.06&  12.7&  7.9&  S&  +&  $-$& 114 \cr
\hline

\end{tabular}

\newpage
\noindent
{\bf Table 1.} Symbiotic stars -- cont\vspace*{0.2cm}.\\
\begin{tabular}{|l|c|c|c|c|c|c|}
\hline \hline
No.& Cool-star spectrum& Radio& {\em IRAS}& {\em IRAS}& References\cr
 & & [mJy]& F$_{12}$[Jy]& F$_{25}$[Jy]& \cr
\hline \hline

151&  M4(\cite{ms99})&  0.28(3.6cm)& & &  \cite{30}(fc,class,spc) \cite{3}(parm)  \cr 
\hline

152&  M3-5(\cite{280})& & & &  \cite{280}(fc,spc,class)  \cr 
\hline

153&  K7/M0(\cite{ms99,16})&  $<$0.22(3.6cm)& & &  \cite{30}(fc,class,spc) \cite{66}(parm)  \cr 
\hline

154&  M3(\cite{3})&  $<$0.3(6cm)& & &  \cite{185}(fc) \cite{1}(spc,class) \cite{3}(parm)  \cr 
\hline

155&  M4,M2(\cite{5,135})&  0.21(3.6cm)& & &  \cite{30}(fc,class,spc) \cite{3}(parm)  \cr 
\hline

156&  M3,M4(\cite{ms99,5})&  0.14(3.6cm)&  0.16&  0.06L&  \cite{30}(fc,class,spc) \cite{336}(parm)  \cr
\hline

\end{tabular}

\newpage
\noindent
{\bf Table 2.} Suspected symbiotic star\vspace*{0.2cm}s\\
\begin{tabular}{|l|c|c|c|c|c|c|c|c|c|c|c|}
\hline \hline
No.& Name& $\alpha$(2000)& $\delta$(2000)& l$^{\rm II}$& b$^{\rm II}$& $V$& $K$& IR& {\em IUE}& X& IP$_{\rm max}$\cr
&&\ $^h$\ \ $^m$\ \ $^s$& \ $^\circ$\ \ '\ \ \ '' &\ $^\circ$&\ $^\circ$& [mag]& [mag]&&&& [eV] \cr
\hline \hline

s01$^\star$&  RAW 1691& 01 18 36.1&  $-$72 42 24.0& & &  15.3&  12& &  &  $-$&  13.6 \cr 
\hline

s02$^\star$&  [BE74] 583& 05 26 54.0&  $-$71 06 00.0& & &  16.1& & &  &  &  13.6 \cr 
\hline

s03$^\star$&  StHA 55& 05 46 42.0&  +06 43 48.0& 199.34& $-$11.12&  13.5& & &  &  &  13.6 \cr 
\hline

s04$^\star$&  GH Gem& 07 04 04.9&  +12 02 12.0& 203.57& +8.23&  14.6&  $>$9.7& &  &  $-$&  \cr 
\hline

s05$^\star$&  ZZ CMi& 07 24 13.9&  +08 53 51.7& 208.64& +11.30&  9.9&  2.8&  S&  &  &  35.1 \cr 
\hline

s06$^\star$&  NQ Gem& 07 31 54.5&  +24 30 12.5& 194.63& +19.35&  7.9&  3.0& &  +&  $-$&  54.4 \cr
\hline

\end{tabular}

\newpage
\noindent
{\bf Table 2.} Suspected symbiotic stars -- cont\vspace*{0.2cm}.\\
\begin{tabular}{|l|c|c|c|c|c|c|}
\hline \hline
No.& Cool-star spectrum& Radio& {\em IRAS}& {\em IRAS}& References\cr
 & & [mJy]& F$_{12}$[Jy]& F$_{25}$[Jy]& \cr
\hline \hline

s01&  C(\cite{199})& & & &  \cite{198}(fc) \cite{199}(spc) \cite{0}(class)  \cr 
\hline

s02&  G-K(\cite{m139})& & & &  \cite{209}(fc) \cite{m139}(class)  \cr 
\hline

s03& C(\cite{61})& & & &  \cite{61}(spc) \cite{0}(class)  \cr 
\hline

s04&  F2:(\cite{178})& & & &  \cite{178}(fc) \cite{31}(class)   \cr 
\hline

s05&  M6,M5(\cite{48,221})& &  6.60&  2.65&  \cite{319}(fc) \cite{48}(spc) \cite{0}(class)  \cr 
\hline

s06&  C6.2(\cite{47})& &  3.28&  0.84&  \cite{226}(spc) \cite{224}(class) \cite{225}(parm)  \cr
\hline

\end{tabular}

\newpage
\noindent
{\bf Table 3.} Orbital photometric ephemerides for symbiotic and suspected
symbiotic star\vspace*{0.2cm}s\\
\begin{tabular}{l l l c l}
\hline
No.& Star & Ephemeris & Eclipse & Ref. \\
\hline
003& EG And & Min(UV) = JD~2445380 + 481$\times$E & Yes & \citetb{337} \\
008& AX Per & Min($\rm m_{vis}/m_{pg}$) = JD~2436667 + 680.8$\times$E & Yes & \citetb{37} \\
017& V1261 Ori &  Min(Vyb) = JD~2445494 + 642$\times$E & Yes & \citetb{213,318} \\
023& BX Mon & Min($\rm m_{pg}$) JD~2449530 + 1401$\times$E & Yes & \citetb{343} \\
024& V694 Mon & Min($\rm m_{pg}$/B) = JD~2437455 + 1930$\times$E & No & \citetb{344} \\
033& SY Mus & Min($\rm m_{vis}$) = JD~2450176 + 625$\times$E & Yes & \citetb{345} \\
045& RW Hya & Min(UV) = JD~2445072 + 370.2$\times$E & Yes & \citetb{m11} \\
\hline
\end{tabular}

\newpage
\noindent
{\bf Table 4.} Orbital elements for symbiotic binaries\vspace*{0.2cm}\\
\begin{tabular}{l l c c c c c l c c l }
\hline
No.& Star & $P$ & $K_g$ & $q \equiv M_{\rm g}/M_{\rm h}$ & $\gamma_0$ & $e$ & T$_{0}$&
$a_g$sin$i$ & $f(M)$ & Ref. \\
   & & [days]& [km/s] & & [km/s] & & [JD] & [R$_\odot$] & [M$_\odot$] & \\
\hline
003& EG And & ~482.6 & ~7.3 & &  ~$-$95 & 0& 2450804$^{(3)}$ & ~70 & 0.020~ & \citetb{373} \\
008& AX Per & ~680.8 & ~7.5 & 2.5 & $-$116.5 & 0 & 2445511.8& 100 & 0.030~ & \citetb{37} \\
011& BD Cam & ~596.21 & ~8.5& & ~$-$22.3& ~0.09& 2442794$^{(1)}$& ~99.7& ~0.037& \citetb{205} \\
017& V1261 Ori & ~642& 7.5& ~3.0& ~79.7& 0.07& 2446778$^{(3)}$& 95& 0.028& \citetb{333} \\
023& BX Mon & 1401 & ~4.3 & 6.7 & ~~29.1 & 0.49 & 2449530& 104 & 0.0076 & \citetb{343} \\
   &        & 1259 & ~4.6 &     & ~~29.1 & 0.44 & 2449680& 103 & 0.0092 & \citetb{373} \\
\hline
\end{tabular}
$^{(1)}$ T$_{0}$ -- time of the passage through periastron\\
$^{(2)}$ assumed from photometric ephemeris (eclipse)\\
$^{(3)}$ time of maximum velocity\\

\newpage
\noindent
{\bf Table 5.} Pulsations ephemerides for Miras in symbiotic and suspected symbiotic
star\vspace*{0.2cm}s\\
\begin{tabular}{l l l c l}
\hline
No.& Star & T$_{0}$ [JD]& P [days]& Ref. \\
\hline
010& o Ceti& Max(V)=2444839& 331.96& \citetb{357}\\
016& UV Aur& Max($\rm m_{pg}$)=2441062& 395.42& \citetb{357}\\
026& RX Pup& Min(J)=2440214& 578& \citetb{Metal99}\\
029& KM Vel& ?& 370& \citetb{m49}\\
030& V366 Car& ?& 433& \citetb{m49}\\
034& BI Cru& ?& 280& \citetb{63}\\
048& V852 Cen& ?& 400& \citetb{m2}\\
049& V835 Cen& ?& 450& \citetb{m49}\\
060& V347 Nor& ?& 370-380& \citetb{22}\\
100& H 1$-$36& ?& 450-500& \citetb{m2}\\
169& HM Sge& Max(J)=2446856& 527& \citetb{m108}\\
174& V1016 Cyg& Min(K)=2444852& 478& \citetb{335}\\
   &      & Max(K)=2445038& 478& \citetb{335}\\
175& RR Tel& Max(J/K)=2442999& 387& \citetb{Fetal83}\\
183& V407 Cyg& Max(B)=2429710& 745& \citetb{163}\\
188& R Aqr& Max(V)=2442398& 386.96& \citetb{357}\\
s25& V850 Aql& Max(V)=2425888& 320& \citetb{357}\\
s28& V627 Cas& ?& 466& \citetb{m192}\\
\hline
\end{tabular}

\newpage
\noindent
{\bf Table 6.} Hipparcos parallaxes for symbiotic stars \vspace*{0.2cm}
\citetb{365}\\
\begin{tabular}{l l r}
\hline
No.& Star& $\pi$\ [mas]\\
\hline
003& EG And& 1.48 $\pm$\ 0.97 \\
010& o Ceti& 7.79 $\pm$\ 1.07 \\
011& BD Cam& 6.27 $\pm$\ 0.63 \\
017& V1261 Ori& 1.32 $\pm$\ 0.99 \\
057& T CrB& $-$1.61 $\pm$\ 0.63 \\
058& AG Dra& $-$1.72 $\pm$\ 0.98 \\
142& AR Pav& 3.37 $\pm$\ 2.47 \\
167& CH Cyg& 3.73 $\pm$\ 0.85 \\
172& CI Cyg& $-$0.36 $\pm$\ 1.58 \\
185& AG Peg& $-$0.30 $\pm$\ 1.17 \\
187& Z And& 2.34 $\pm$\ 2.91 \\
188& R Aqr& 5.07 $\pm$\ 3.15 \\
\hline
\end{tabular}

\newpage    
\noindent   
{\bf Table 7.} Flickering and outburst characteristics of symbiotic 
and suspected symbiotic stars. SyN -- symbiotic nova, SyRN -- 
symbiotic recurrent nova, Z And -- Z And type outburs\vspace*{0.2cm}t.\\
\begin{tabular}{l l c r c c l}
\hline
No.& Star& Flick.& Ref.& Out.& Type of Out.& Ref. \\
\hline
002 & SMC2& & & +& & \citetb{34}\\
004 & SMC3& & & +& & \citetb{34}\\
008 & AX Per& & & +& Z And& \citetb{37}\\
010 & o Ceti& +& \citetb{m236}& & & \\
012 & S32& & & +?& & \citetb{m29}\\
013 & LMC S154& & & +?& & \citetb{208}\\
024 & V694 Mon& +& \citetb{m112}& +& & \citetb{293}\\
\hline
\end{tabular}

\newpage
\noindent
{\bf Table 8.} Different names for symbiotic and suspected symbiotic star\vspace*{0.2cm}s\\
001=SMC 1=NAME SMC1=[MH95] 183
\vspace*{-0.3cm}\\

\noindent
002=SMC 2=NAME SMC2
\vspace*{-0.3cm}\\

\noindent
003=EG And=HD 4174=BD+39 167=SAO 36618=GCRV 403=HIC 3494=GEN\# +1.00004174=
AG+40 66=GC 880=DO 8473=GPM1 20=SKY\# 1157=AGKR 609=IRC +40014=JP11 413=
PPM 43262=HIP 3494={\em IRAS} 00415+4024
\vspace*{-0.3cm}\\

\noindent
004=SMC 3=NAME SMC3=RX J0048.4$-$7332
\vspace*{-0.3cm}\\

\noindent
005=SMC N60=LHA 115$-$N 60=LIN 323=HV 1707(???)
\vspace*{-0.3cm}\\

\noindent
007=SMC N73=LHA 115$-$N 73=LIN 445 a
\vspace*{-0.3cm}\\

\noindent
008=AX Per=MWC 411=HV 5488=CSI+54$-$01331=GCRV 896=JP11 5465={\em IRAS} 01331+5359
\vspace*{-0.3cm}\\

\noindent
009=V471 Per=PN M 1$-$2=PK 133$-$08 1=PN VV 8=LS V +52 1=PN G133.1$-$08.6=PN ARO
116=CSI+52$-$01555=PN VV' 11={\em IRAS} 01555+5239
\vspace*{-0.3cm}\\

\noindent
010=o Cet=MIRA=HD 14386=RAFGL 318=SKY\# 3428=GC 2796=omi Cet=ADS 1778 AP=
IRC +00030=YZ 93 562=MWC 35=BD$-$03 353=GEN\# +1.00014386J=
CCDM J02194$-$0258AP=PLX 477=GCRV 1301=JP11 625=CSI$-$03 353 1=DO 430=
HIC 10826=SAO 129825=68 Cet=HR 681=UBV 21604=LTT 1179=TD1 1361=
PPM 184482=JOY 1AP={\em IRAS} 02168$-$0312
\vspace*{-0.3cm}\\

\noindent
011=BD Cam=HD 22649=BD+62 597=HR 1105=SAO 12874=GC 4383=IRC +60125=FK4 129=   
UBV 3468=CSS 79=[HFE83] 244=GEN\# +1.00022649=AG+63 277=N30 751=
GCRV 2027=RAFGL 506=PPM 14446=HIP 17296=SKY\# 5606=UBV M 9615=PLX 758=
JP11 803=CSV 328=HIC 17296=S1* 60={\em IRAS} 03377+6303
\vspace*{-0.3cm}\\

\noindent
012=S32=StHA 32
\vspace*{-0.3cm}\\

\noindent
013=LMC S154=LHA 120$-$S 154
\vspace*{-0.3cm}\\

\noindent
014=LMC S147=LHA 120$-$S 147=[BE74] 484
\vspace*{-0.3cm}\\

\noindent
015=LMC N19=LHA 120$-$N 19=[BE74] 191=LMC B0503$-$6803=MC4(0503$-$680)=
PMN J0503$-$6757
\vspace*{-0.3cm}\\

\noindent
016=UV Aur=HD 34842=BD+32 957=ADS 3934 A=IRC +30110=SAO 57941=HV 3322=Case 9=
GEN\# +1.0003482A=CGCS 911=IDS 05153+3224 A=AG+32 505=JP11 1034=UBV M 10852=
AN 58.1911=LEE 179=RAFGL 735=PPM 70251=HIC 25050=GCRV 3199=
DO 11210=CSI+32 957 1=Fuen C 29=CCDM J05218+3231A=C* 318={\em IRAS} 05185+3227
\vspace*{-0.3cm}\\

\noindent
017=V1261 Ori=HD 35155=BD$-$08 1099=RAFGL 736=SAO 132035=GCRV 56103=YZ 98
1455=
HIC 25092=GEN\# +1.00035155=GC 6602=HERZ 11764=CSS 133=HIP 25092=SKY\# 8520=
IRC $-$10086=PPM 187990=S1* 98={\em IRAS} 05199$-$0842
\vspace*{-0.3cm}\\

\noindent
019=LMC N67=CH$-$95=LHA 120$-$N 67=[HC88] 95
\vspace*{-0.3cm}\\

\noindent
021=LMC S63=LHA 120$-$S 63=HV 12671
\vspace*{-0.3cm}\\

\noindent
022=SMP LMC 94=LHA 120$-$S 170=LM 2$-$44
\vspace*{-0.3cm}\\

\noindent
023=BX Mon=AS 150=HV 10446=MHA 61$-$12=JP11 5448=SS73 4
\vspace*{-0.3cm}\\

\noindent
024=V694 Mon=MWC 560=SS73 5=LS 391=CSI$-$07$-$07234=GSC 05396$-$01135=
MHA 61$-$14={\em IRAS} 07233$-$0737
\vspace*{-0.3cm}\\

\noindent
025=WRAY 15$-$157={\em IRAS} 08045$-$2823
\vspace*{-0.3cm}\\

\noindent
026=RX Pup=HD 69190=WRAY 16$-$17=HV 3372=CPD$-$41 2287=Hen 3$-$138=AN 88.1914=SS73 8=
JP11 1653=PK 258$-$03 1=CD$-$41 3911={\em IRAS} 08124$-$4133
\vspace*{-0.3cm}\\

\noindent
027=Hen 3$-$160=SS73 9=WRAY 15$-$208
\vspace*{-0.3cm}\\

\noindent
028=AS 201=MHA 382$-$43=Hen 3$-$172=SCM 38=PK 249+06 1=PDS 32=PN SaSt 1$-$1=
PN G249.0+06.9={\em IRAS} 08296$-$2735
\vspace*{-0.3cm}\\

\noindent
029=KM Vel=Hen 2$-$34=WRAY 16$-$56=PK 274+02 1=ESO 212$-$13=PN G274.1+02.5=
SS73 14={\em IRAS} 09394$-$4909
\vspace*{-0.3cm}\\

\noindent
030=V366 Car=Hen 2$-$38=PK 280$-$02 1=PN SaSt 1$-$3=ESO 167$-$11=WRAY 16$-$62=
{\em IRAS} 09530$-$5704
\vspace*{-0.3cm}\\

\noindent
031=Hen 3$-$461=PK 283+06 2={\em IRAS} 10370$-$5108
\vspace*{-0.3cm}\\

\noindent
033=SY Mus=HD 100336=SS73 32=HV 3376=WRAY 15$-$824=TD1 15706=Hen 3$-$667=
AN 118.1914=CPD$-$65 11298=CSI$-$65$-$11298=SPH 127={\em IRAS} 11299$-$6508
\vspace*{-0.3cm}\\

\noindent
034=BI Cru=Hen 3$-$782=SVS 1855={\em IRAS} 12206$-$6221
\vspace*{-0.3cm}\\

\noindent
035=RT Cru=HV 1245=AN 131.1906
\vspace*{-0.3cm}\\

\noindent
036=TX CVn=BD+37 2318=UBV M 2779=PPM 76696=JP11 5415=CDS 866=FB 110=
Case A-F 788=AG+37 1208=SAO 63173=GEN\# +0.03702318={\em IRAS} 12423+3702
\vspace*{-0.3cm}\\

\noindent
037=Hen 2$-$87=SS73 36=PK 302$-$00 1=WRAY 16$-$119=ESO 95$-$16={\em IRAS} 12423$-$6244
\vspace*{-0.3cm}\\

\noindent
038=Hen 3$-$828=SS73 37=WRAY 15$-$1022
\vspace*{-0.3cm}\\

\noindent
039=SS73 38=CD$-$64 665=CGCS 3295=DJM 390={\em IRAS} 12483$-$6443
\vspace*{-0.3cm}\\

\noindent
041=St 2$-$22=PN Sa 3$-$22=PK 305+03 1
\vspace*{-0.3cm}\\

\noindent
042=CD$-$36 8436=NSV 06160=Hen 3$-$886=JP11 311=PK 308+25 6={\em IRAS} 13131$-$3644
\vspace*{-0.3cm}\\

\noindent
043=V840 Cen=LILLER'S OBJECT
\vspace*{-0.3cm}\\

\noindent
044=Hen 3$-$905=SS73 40=WRAY 15$-$1108
\vspace*{-0.3cm}\\

\noindent
045=RW Hya=HD 117970=MWC 412=CPD$-$24 5101=YZ 115 9978=SAO 181760=CD$-$24 10977=
GCRV 8034=HV 3237=PPM 261808=GEN\# +1.00117970=JP11 2404=AN 51.1910=
FAUST 3820={\em IRAS} 13315$-$2507
\vspace*{-0.3cm}\\

\noindent
046=Hen 3$-$916=SS73$-$42=WRAY 15$-$1123
\vspace*{-0.3cm}\\

\noindent
047=V704 Cen=Hen 2$-$101=PK 311+03 1=PN G311.1+03.4=WRAY 16$-$141=ESO 133$-$7=
{\em IRAS} 13515$-$5812
\vspace*{-0.3cm}\\

\noindent
048=V852 Cen=Hen 2$-$104=PK 315+09 1=ESO 221$-$31=PN G315.4+09.4=SS73 43=
NAME SOUTHERN CRAB=WRAY 16$-$147=SCM 78={\em IRAS} 14085$-$5112
\vspace*{-0.3cm}\\

\noindent
049=V835 Cen=Hen 2$-$106=WRAY 16$-$148=PK 312$-$02 1=ESO 97$-$14=SCM 79=NSV 06587=
{\em IRAS} 14103$-$6311
\vspace*{-0.3cm}\\

\noindent
050=V417 Cen=HV 6516={\em IRAS} 14122$-$6139
\vspace*{-0.3cm}\\

\noindent
052=Hen 2$-$127=SS73 45=PK 325+04 2=ESO 224$-$2=SCM 94=WRAY 16$-$180=
{\em IRAS} 15210$-$5139
\vspace*{-0.3cm}\\

\noindent
053=Hen 3$-$1092=Hen 2$-$134=SS73 46=PK 319$-$09 1=WRAY 16$-$186=LS 3391=ESO 99$-$10=
JP11 5230=CSI$-$66$-$15425
\vspace*{-0.3cm}\\

\noindent
054=Hen 3$-$1103=SS73 47=WRAY 15$-$1359
\vspace*{-0.3cm}\\

\noindent
055=HD 330036=BD$-$48 10371=PN Cn 1$-$1=PK 330+04 1=CD$-$48 10371=HIP 77662=
ESO 225$-$1=HIC 77662=WRAY 15$-$1364=PN G330.7+04.1={\em IRAS} 15476$-$4836 
\vspace*{-0.3cm}\\

\noindent
056=Hen 2$-$139=PK 326$-$01 1=WRAY 16$-$193=ESO 178$-$2={\em IRAS} 15508$-$5520
\vspace*{-0.3cm}\\

\noindent
057=T CrB=HD 143454=MWC 413=GC 21491=IDS 15553+2612 AB=DO 15377=PPM 104498=
GEN\# +1.00143454J=BD+26 2765=GCRV 9203=SAO 84129=NOVA CrB 1866=
CCDM J15595+2555AB=AG+26 1536=HR 5958=SBC 558=NOVA CrB 1946=HIC 78322= 
HIP 78322={\em IRAS} 15574+2603
\vspace*{-0.3cm}\\

\noindent
058=AG Dra=AG+66 715=CDS 889=SAO 16931=HIC 78512=1ES 1601+66.9=
BPS BS 16087$-$0012=BD+67 922=GCRV 9231=UBV 13635=GEN\# +0.06700922=
2E 1601.3+6656=2RE J1601+664=HIP 78512=JP11 236=SVS 1155=PPM 19692=
2E 3573=2RE J160133+664805={\em IRAS} 16013+6656
\vspace*{-0.3cm}\\

\noindent
059=WRAY 16$-$202=PN Sa 3$-$33=PK 332+01 1
\vspace*{-0.3cm}\\

\noindent
060=V347 Nor=Hen 2$-$147=WRAY 16$-$208=PK 327$-$04 1=ESO 178$-$13=PN G327.9$-$04.3=
{\em IRAS} 16099$-$5651
\vspace*{-0.3cm}\\

\noindent
061=UKS Ce$-$1=UKS$-$Ce1=UKS Ce1=UKS 1612$-$22.0
\vspace*{-0.3cm}\\

\noindent
062=QS Nor=Hen 2$-$156=SS73 53=WRAY 15$-$1461=PK 338+05 2=SCM 111=ESO 331$-$2=
CSV 2635
\vspace*{-0.3cm}\\

\noindent
063=WRAY 15$-$1470=Hen 3$-$1187=SS73 55
\vspace*{-0.3cm}\\

\noindent
064=Hen 2$-$171=SS73 59=WRAY 16$-$226=PK 346+08 1=PN G346.0+08.5=ESO 390$-$7=
{\em IRAS} 16307$-$3459
\vspace*{-0.3cm}\\

\noindent
065=Hen 3$-$1213=SS73 60=WRAY 15$-$1511
\vspace*{-0.3cm}\\

\noindent
066=Hen 2$-$173=SS73 61=WRAY 15$-$1518=PK 342+05 1=ESO 331$-$7
\vspace*{-0.3cm}\\

\noindent
067=Hen 2$-$176=PK 339+00 1=WRAY 16$-$230=ESO 277$-$5={\em IRAS} 16379$-$4507
\vspace*{-0.3cm}\\

\noindent
068=KX TrA=Hen 3$-$1242=PN Cn 1$-$2=Hen 2$-$177=PK 326$-$10 1=WRAY 16$-$233=
ESO 101$-$10=JP11 5232=PN StWr 1$-$1=JP11 5233={\em IRAS} 16401$-$6231
\vspace*{-0.3cm}\\

\noindent
069=AS 210=MHA 276$-$52=Hen 3$-$1265=SS73 66=JP11 5249=WRAY 16$-$237=PK 355+11 1=
{\em IRAS} 16482$-$2555
\vspace*{-0.3cm}\\

\noindent
070=HK Sco=HV 4493=AS 212=MHA 71$-$6=Hen 3$-$1280=SS73 68=WRAY 15$-$1563
\vspace*{-0.3cm}\\

\noindent
071=CL Sco=HV 4035=AS 213=MHA 71$-$5=Hen 3$-$1286=SS73 69=WRAY 15$-$1564=
CD$-$30 13603
\vspace*{-0.3cm}\\

\noindent
072=PN MaC 1$-$3=PK 339$-$03 1
\vspace*{-0.3cm}\\

\noindent
073=V455 Sco=HV 7869=AS 217=MHA 71$-$15=Hen 3$-$1334=SS73 74=PN H 2$-$2=
PK 351+03 1=ESO 392$-$3=WRAY 16$-$252=Haro 3$-$2=PN VV' 162={\em IRAS} 17040$-$3401
\vspace*{-0.3cm}\\

\noindent
074=Hen 3$-$1341=NSV 08226=SS73 75
\vspace*{-0.3cm}\\

\noindent
075=Hen 3$-$1342=SS73 77
\vspace*{-0.3cm}\\

\noindent
076=AS 221=MHA 276$-$12=Hen 3$-$1348=SS73 79=PN H 2$-$4=PK 352+03 1=WRAY 15$-$1637=
PN VV' 166=Haro 3$-$4=ESO 392$-$4={\em IRAS} 17087$-$3230
\vspace*{-0.3cm}\\

\noindent
077=PN H 2$-$5=PK 354+04 2=PN VV' 172=ESO 454$-$3=WRAY 15$-$1655=Haro 3$-$5=
{\em IRAS} 17121$-$3131
\vspace*{-0.3cm}\\

\noindent
078=PN Sa 3$-$43=PK 355+04 1
\vspace*{-0.3cm}\\

\noindent
079=Draco C$-$1=Irwin Dra 22025=Stetson 3203=McClure C1=[ALS82] C1
\vspace*{-0.3cm}\\

\noindent
080=PN Th 3$-$7=PK 356+04 3=PN ARO 249=ESO 454$-$12
\vspace*{-0.3cm}\\

\noindent
081=PN Th 3$-$17=PK 357+03 3=WRAY 17$-$81=ESO 454$-$30=PN Sa 3$-$54
\vspace*{-0.3cm}\\

\noindent
082=PN Th 3$-$18=PK 358+03 5=ESO 454$-$32=WRAY 17$-$83=Hen 2$-$228
\vspace*{-0.3cm}\\

\noindent
083=Hen 3$-$1410=PN Th 3$-$20=PK 357+02 3=PN ARO 251=NSV 08805=WRAY 15$-$1714=
ESO 454$-$37
\vspace*{-0.3cm}\\

\noindent
084=V2116 Oph=GX 1+04=PK 001+04 1=4U 1728$-$24=1M 1728$-$247=H 1728$-$247=
PN MaC 1$-$5=2S 1728$-$247=GX 2+05=2U 1728$-$24=3U 1728$-$24=3A 1728$-$247=
1H 1728$-$247
\vspace*{-0.3cm}\\

\noindent
085=PN Th 3$-$29=WRAY 17$-$89=PN Sa 3$-$61=PK 358+02 3=ESO 455$-$15=
{\em IRAS} 17299$-$2905
\vspace*{-0.3cm}\\

\noindent
086=PN Th 3$-$30=PK 359+02 1=ESO 455$-$18=WRAY 17$-$90=Hen 2$-$243=PN Sa 3$-$63
\vspace*{-0.3cm}\\

\noindent
087=PN Th 3$-$31=PK 358+01 2=ESO 455$-$19=Hen 2$-$245=WRAY 17$-$91=
{\em IRAS} 17312$-$2926
\vspace*{-0.3cm}\\

\noindent
088=PN M 1$-$21=SS73 90=Hen 2$-$247=PK 006+07 1=PN VV' 216=PN VV 103=
ESO 588$-$7={\em IRAS} 17313$-$1909
\vspace*{-0.3cm}\\

\noindent
089=Hen 2$-$251=PK 358+01 3=PN H 1$-$25=PN VV' 218=Haro 2$-$25=WRAY 16$-$289=
SS73 92=SCM 151=ESO 455$-$22=PN Bl A={\em IRAS} 17321$-$2943
\vspace*{-0.3cm}\\

\noindent
090=PN Pt 1=PK 003+03 3=EQ 1735$-$2352=ESO 520$-$12
\vspace*{-0.3cm}\\

\noindent
091=PN K 6$-$6=WR 99=LuSt 1=PN G359.8+01.5=Terz V 2955  
\vspace*{-0.3cm}\\

\noindent
092=RT Ser=MWC 265=SS73 94=CSI$-$11$-$17371=NOVA Ser 1909=NOVA Ser 1917=
GCRV 10203=AN 7.1917
\vspace*{-0.3cm}\\

\noindent
093=AE Ara=HV 5491=MWC 591=Hen 3$-$1451=SS73 95=WRAY 15$-$1754=PN PC 18=
PK 344$-$08 1=GSC 08347$-$01978=ESO 279$-$5
\vspace*{-0.3cm}\\

\noindent
095=UU Ser=AS 237=SS73 98=AN 720.1936=HV 8771=CSV 2420
\vspace*{-0.3cm}\\

\noindent
096=V2110 Oph=AS 239=MHA 79$-$52=Hen 3$-$1465=JP11 5250
\vspace*{-0.3cm}\\

\noindent
097=V916 Sco=SSM 1=EQ 1740$-$360
\vspace*{-0.3cm}\\

\noindent
098=Hen 2$-$275=WRAY 16$-$304=PK 351$-$05 1=ESO 334$-$8
\vspace*{-0.3cm}\\

\noindent
099=V917 Sco=Hen 3$-$1481=SS73 103
\vspace*{-0.3cm}\\

\noindent
100=PN H 1$-$36=PK 353$-$04 1=Hen 2$-$289=PN Sa 2$-$249=PN G353.5$-$04.9=
ESO 393$-$31=PN VV' 259=Haro 2$-$36={\em IRAS} 17463$-$3700
\vspace*{-0.3cm}\\

\noindent
101=RS Oph=HD 162214=MWC 414=BD$-$06 4661=NOVA Oph 1898=AN 20.1901=
PPM 201101=GCRV 10316=HV 164=SS73 106=JP11 2898=NOVA Oph 1933=
{\em IRAS} 17474$-$0641
\vspace*{-0.3cm}\\

\noindent
102=WRAY 16$-$312=PN Sa 3$-$80=PK 358$-$01 2
\vspace*{-0.3cm}\\

\noindent
103=V4141 Sgr=PN Th 4$-$4=PK 008+03 2=ESO 589$-$10=PN ARO 260={\em IRAS} 17477$-$1948
\vspace*{-0.3cm}\\

\noindent
104=ALS 2=[ALS88] 2=SS 324=EQ 174755.9$-$174710=PK 010+04 7
\vspace*{-0.3cm}\\

\noindent
105=AS 245=MHA 359$-$110=Hen 3$-$1501=SS73 107=PN H 2$-$28=PK 006+02 2=
ESO 589$-$13=PN VV' 268=PN ARO 261=Haro 3$-$28={\em IRAS} 17479$-$2218
\vspace*{-0.3cm}\\

\noindent
106=Hen 2$-$294=PK 357$-$03 1=ESO 394$-$1=WRAY 16$-$318={\em IRAS} 17483$-$3250
\vspace*{-0.3cm}\\

\noindent
107=PN Bl 3$-$14=PK 000$-$01 4=ESO 455$-$55
\vspace*{-0.3cm}\\

\noindent
108=PN Bl 3$-$6=PK 358$-$02 2=PN Sa 3$-$85=ESO 456$-$4={\em IRAS} 17496$-$3120
\vspace*{-0.3cm}\\

\noindent
109=PN Bl L=PK 359$-$02 1=PN Sa 3$-$86=ESO 456$-$6
\vspace*{-0.3cm}\\

\noindent
110=V745 Sco=NOVA Sco 1937
\vspace*{-0.3cm}\\

\noindent
111=PN MaC 1$-$9=PK 013+05 2
\vspace*{-0.3cm}\\

\noindent
112=AS 255=MHA 363$-$45=Hen 3$-$1525=SS73 111={\em IRAS} 17537$-$3513
\vspace*{-0.3cm}\\

\noindent
113=V2416 Sgr=PN M 3$-$18=SS73 112=PK 007+01 2=ESO 589$-$24=PN VV' 295=
Hen 2$-$312=Ve 2$-$61={\em IRAS} 17542$-$2142
\vspace*{-0.3cm}\\

\noindent
114=PN H 2$-$34=PN Sa 3$-$105=PK 001$-$02 1=ESO 456$-$28=PN VV' 304=Haro 3$-$34
\vspace*{-0.3cm}\\

\noindent
116=AS 269=PN H 1$-$49=SS73 119=PK 358$-$05 2=Hen 2$-$331=MHA 304$-$52=ESO 394$-$23=
WRAY 16$-$356=Haro 2$-$49=PN VV' 323=SCM 178={\em IRAS} 18001$-$3242
\vspace*{-0.3cm}\\

\noindent
117=PN Ap 1$-$8=SS73 121=PK 002$-$03 1=ESO 456$-$47={\em IRAS} 18013$-$2821
\vspace*{-0.3cm}\\

\noindent
118=SS73 122=PN KFL 6={\em IRAS} 18015$-$2709
\vspace*{-0.3cm}\\

\noindent
119=AS 270=Hen 3$-$1581=SS73 126={\em IRAS} 18026$-$2025
\vspace*{-0.3cm}\\

\noindent
120=PN H 2$-$38=SS73 128=PK 002$-$03 4=Hen 2$-$343=WRAY 17$-$108=ESO 456$-$51=
PN VV' 342=Haro 3$-$38={\em IRAS} 18028$-$2817
\vspace*{-0.3cm}\\

\noindent
121=SS73 129=PK 001$-$04 3=T 17=PN KFL 8={\em IRAS} 18038$-$2932
\vspace*{-0.3cm}\\

\noindent
122=Hen 3$-$1591=SS73 132=T 53=NSV 10219={\em IRAS} 18044$-$2558
\vspace*{-0.3cm}\\

\noindent
123=V615 Sgr=Hen 2$-$349=PK 356$-$07 1=SS73 131=ESO 394$-$29=WRAY 15$-$1840=
HV 7199={\em IRAS} 18044$-$3610
\vspace*{-0.3cm}\\

\noindent
124=Ve 2$-$57=SS 134=NSV 10241
\vspace*{-0.3cm}\\

\noindent
125=AS 276=Hen 3$-$1595=SS73 135=MHA 363$-$7={\em IRAS} 18058$-$4108
\vspace*{-0.3cm}\\

\noindent
126=PN Ap 1$-$9=SS73 137=PK 003$-$04 2=Hen 2$-$356=WRAY 16$-$377=ESO 456$-$63=
SCM 186={\em IRAS} 18073$-$2753
\vspace*{-0.3cm}\\

\noindent
127=AS 281=MHA 208$-$83=SS73 138=PN Ap 1$-$10=PK 003$-$04 1=WRAY 16$-$378=
ESO 456$-$65=Hen 2$-$357={\em IRAS} 18076$-$2757
\vspace*{-0.3cm}\\

\noindent
128=V2506 Sgr=AS 282=MHA 304$-$113=Hen 2$-$358=SS73 139=PN Ap 1$-$11=
PK 003$-$04 6=WRAY 16$-$379=ESO 456$-$66
\vspace*{-0.3cm}\\

\noindent
129=SS73 141=PK 359$-$07 2=WRAY 16$-$384
\vspace*{-0.3cm}\\

\noindent
130=AS 289=Hen 3$-$1627=SS73 143=F1$-$11={\em IRAS} 18095$-$1140
\vspace*{-0.3cm}\\

\noindent
131=Y CrA=HD 166813=Hen 3$-$1632=CSI$-$42$-$18107=SS73 144=HV 169=
PK 350$-$11 1=AN 25.1901={\em IRAS} 18110$-$4252
\vspace*{-0.3cm}\\

\noindent
132=YY Her=AS 297=CSI+20$-$18124=GSC 01579$-$00381=AN 6.1919=JP11 5444=
MHA 352$-$34
\vspace*{-0.3cm}\\

\noindent
133=V2756 Sgr=AS 293=MHA 304$-$122=Hen 2$-$370=SS73 145=PK 002$-$05 1=ESO 456$-$79=
WRAY 16$-$392
\vspace*{-0.3cm}\\

\noindent
134=FG Ser=AS 296=D 143$-$2=SS73 148=SON 10363=JP11 5251={\em IRAS} 18125$-$0019
\vspace*{-0.3cm}\\

\noindent
135=HD 319167=PN CnMy 17=Hen 2$-$373=SS73 146=PK 001$-$06 1=WRAY 16$-$395=
SCM 195=ESO 456$-$81
\vspace*{-0.3cm}\\

\noindent
136=Hen 2$-$374=SS73 147=PK 009$-$02 1=ESO 590$-$10={\em IRAS} 18126$-$2135
\vspace*{-0.3cm}\\

\noindent
137=Hen 2$-$376=AS 294=NSV 10435=PK 004$-$05 2=MHA 304$-$123=WRAY 16$-$396=
PN Sa 3$-$126=ESO 457$-$1
\vspace*{-0.3cm}\\

\noindent
138=V4074 Sgr=AS 295B=Hen 3$-$1641=HIC 89526=HIP 89526
\vspace*{-0.3cm}\\

\noindent
139=V2905 Sgr=AS 299=SS73 151=GEN\# +6.20010299=MHA 208$-$92
\vspace*{-0.3cm}\\

\noindent
141=Hen 3$-$1674=SS73 153=T 21=WRAY 15$-$1864=PN KFL 17
\vspace*{-0.3cm}\\

\noindent
142=AR Pav=MWC 600=CPD$-$66 3307=GCRV 10756=HIC 89886=HV 7860=PPM 363277=
HIP 89886=SBC 668=GSC 09080$-$00788={\em IRAS} 18157$-$6609
\vspace*{-0.3cm}\\

\noindent
143=V3929 Sgr=Hen 2$-$390=SS73 154=CSV 4026=PK 005$-$05 2=ESO 522$-$19=
PN ARO 274=PN StWr 2$-$3=SCM 202=HV 9397=P 4629={\em IRAS} 18178$-$2649
\vspace*{-0.3cm}\\

\noindent
144=V3804 Sgr=AS 302=MHA 304$-$33=Hen 3$-$1676=SS73 155={\em IRAS} 18182$-$3133
\vspace*{-0.3cm}\\

\noindent
145=V443 Her=MWC 603=GCRV 68111=CSI+23$-$18201=JP11 5216=FB 171=
{\em IRAS} 18200+2325
\vspace*{-0.3cm}\\

\noindent
146=V3811 Sgr=Hen 2$-$396=ESO 590$-$19=PK 010$-$03 1=SS73 160={\em IRAS} 18206$-$2157
\vspace*{-0.3cm}\\

\noindent
147=V4018 Sgr=CD$-$28 14567=AS 304=Hen 3$-$1691=SS73 162=PN KFL 20=
GSC 06869$-$00806={\em IRAS} 18221$-$2837
\vspace*{-0.3cm}\\

\noindent
148=V3890 Sgr=NOVA Sgr 1962=SS 390
\vspace*{-0.3cm}\\

\noindent
149=V2601 Sgr=AS 313=MHA 208$-$51=SS73 171={\em IRAS} 18349$-$2244
\vspace*{-0.3cm}\\

\noindent
150=PN K 3$-$9=PN Sa 3$-$142=PK 023$-$01 1={\em IRAS} 18376$-$0846
\vspace*{-0.3cm}\\

\noindent
151=AS 316=MHA 208$-$58=Hen 2$-$417=SS73 172=PK 012$-$07 1=ESO 591$-$14
\vspace*{-0.3cm}\\

\noindent
152=DQ Ser=CSV 4342
\vspace*{-0.3cm}\\

\noindent
153=MWC 960=MHA 204$-$22=Hen 3$-$1726=SS73 174
\vspace*{-0.3cm}\\

\noindent
154=AS 323=PN K 4$-$7=PK 026$-$02 2=MHA 369$-$39=PN ARO 292
\vspace*{-0.3cm}\\

\noindent
155=AS 327=MHA 208$-$67=Hen 3$-$1730=SS73 176=PK 011$-$11 1=JP11 5253=WRAY 16$-$421
\vspace*{-0.3cm}\\

\noindent
156=FN Sgr=AS 329=SS73 177=CSI$-$19$-$18509=NOVA Sgr 1925=GCRV 11342=JP11 5475
\vspace*{-0.3cm}\\

\noindent
157=PN Pe 2$-$16=PK 029$-$02 1=PN Th 1$-$F=PN ARO 296=PN VV' 464
\vspace*{-0.3cm}\\

\noindent
159=V919 Sgr=AS 337=MHA 227$-$6=SS73 178=AN 237.1932
\vspace*{-0.3cm}\\

\noindent
160=V1413 Aql=AS 338=MHA 305$-$6=Hen 3$-$1737=PN K 4$-$12=PK 048+04 1=SS 428=
{\em IRAS} 19015+1625
\vspace*{-0.3cm}\\

\noindent
162=PN Ap 3$-$1=PK 037$-$02 1=SCM 231=PN ARO 145
\vspace*{-0.3cm}\\

\noindent
163=PN MaC 1$-$17=PK 030$-$07 1
\vspace*{-0.3cm}\\

\noindent
164=V352 Aql=PN K 3$-$25=PK 037$-$03 3=PN ARO 306=AN 279.1931={\em IRAS} 19111+0212
\vspace*{-0.3cm}\\

\noindent
165=ALS 1=[ALS88] 1=EQ 191333.1$-$082308=PK 028$-$09 3
\vspace*{-0.3cm}\\

\noindent
166=BF Cyg=MWC 315=LS II +29 5=JP11 5433=CSI+29$-$19219 2=GCRV 11847=
AN 112.1914
\vspace*{-0.3cm}\\

\noindent
167=CH Cyg=HD 182917=SAO 31632=BD+49 2999=GCRV 11865=JP11 3103=RAFGL 2383=
HIC 95413=GEN\# +1.00182917=AG+50 1370=DO 37228=YZ 50 6001=HIP 95413=
SKY\# 36122=GC 26820=CDS 1064=IRC +50294=AN 24.1924=PPM 37375=
{\em IRAS} 19232+5008
\vspace*{-0.3cm}\\

\noindent
169=HM Sge=PK 053$-$03 2=2E 1939.7+1637=SVS 2183=2E 4280={\em IRAS} 19396+1637
\vspace*{-0.3cm}\\

\noindent
170=Hen 3$-$1761=NSV 12264={\em IRAS} 19373$-$6814
\vspace*{-0.3cm}\\

\noindent
171=QW Sge=AS 360=MHA 80$-$5=Hen 3$-$1771=JP11 5254=NSV 12383
\vspace*{-0.3cm}\\

\noindent
172=CI Cyg=MWC 415=HV 3625=CSI+35$-$19484=JP11 5434=HIP 97594=GCRV 12195=
AN 10.1922=HIC 97594={\em IRAS} 19483+3533
\vspace*{-0.3cm}\\

\noindent
173=StHA 169=NSV 12466
\vspace*{-0.3cm}\\

\noindent
174=V1016 Cyg=AS 373=CSI+39$-$19553=JP11 5437=2E 1955.3+3941=PK 075+05 1=
JP11 5438=2E 4302=GCRV 70112=MHA 328$-$116={\em IRAS} 19553+3941
\vspace*{-0.3cm}\\

\noindent
175=RR Tel=Hen 3$-$1811=SKY\# 37701=AN 166.1908=2E 4313=CSI$-$55$-$20003=
GCRV 6924 E=HV 3181=2E 2000.3$-$5552={\em IRAS} 20003$-$5552
\vspace*{-0.3cm}\\

\noindent
176=PU Vul=NOVA Vul 1979=KUWANO-HONDA={\em IRAS} 20189+2124
\vspace*{-0.3cm}\\

\noindent
177=LT Del=Hen 2$-$467=PK 063$-$12 1=PN ARO 353=StHA 179
\vspace*{-0.3cm}\\

\noindent
179=Hen 2$-$468=PK 075$-$04 1=PN ARO 355
\vspace*{-0.3cm}\\

\noindent
180=ER Del=BD+08 4506=AG+08 2842=SVS 654
\vspace*{-0.3cm}\\

\noindent
181=V1329 Cyg=HBV 475=PK 077$-$05 1=UBV M 46710=VES 248={\em IRAS} 20492+3518
\vspace*{-0.3cm}\\

\noindent
182=CD$-$43 14304=Hen 3$-$1924
\vspace*{-0.3cm}\\

\noindent
183=V407 Cyg=AS 453=MHA 289$-$90=AN 148.1940=NOVA Cyg 1936
\vspace*{-0.3cm}\\

\noindent
184=StHA 190=RJHA 120={\em IRAS} 21392+0230
\vspace*{-0.3cm}\\

\noindent
185=AG Peg=HD 207757=MWC 379=AGKR 19529=JP11 3412=TD1 28516=SKY\# 41636=
HIC 107848=GEN\# +1.00207757=BD+11 4673=SAO 107436=DO 7622=HIP 107848=
AG+12 2570=PLX 5279=GCRV 13724=YZ 12 8693=SBC 879=PPM 140717=
{\em IRAS} 21486+1223
\vspace*{-0.3cm}\\

\noindent
186=LL Cas=PN K 4$-$46=PK 108$-$05 1=AN 207.1940=PN ARO 379
\vspace*{-0.3cm}\\

\noindent
187=Z And=HD 221650=MWC 416=PLX 5697=SAO 53146=2E 2331.6+4834=GEN\#
+1.00221650=
BD+48 4093=GCRV 14773=HV 193=PPM 64386=2E 4735=AG+48 2087=JP11 3636=
AN 41.1901=HIC 116287=HIP 116287={\em IRAS} 23312+4832
\vspace*{-0.3cm}\\

\noindent
188=R Aqr=HD 222800=SAO 165849=HR 8992=GCRV 14862=MWC 400=SKY\# 44830=
GC 32948=YZ 105 8733=PPM 242022=HIP 117054=GEN\# +1.00222800=IRC $-$20642=
BD$-$16 6352=RAFGL 3136=HIC 117054={\em IRAS} 23412$-$1533
\vspace*{-0.3cm}\\

\noindent
s01=RAW 1691=[MA93] 1858=LIN 521
\vspace*{-0.3cm}\\

\noindent
s04=GH Gem=CSV 950=AN 241.1943
\vspace*{-0.3cm}\\ 

\noindent
s05=ZZ CMi=BD+09 1633=IRC +10162=AN 306.1934=GCRV 4915=HIP 35915=HIC 35915=
{\em IRAS} 07214+0859
\vspace*{-0.3cm}\\ 

\noindent
s06=NQ Gem=HD 59643=BD+24 1686=SAO 79474=N30 1687=DO 13087=YZ 24 2992=
GEN\# +1.00059643=AG+24 848=YZ 0 823=PPM 97628=CGCS 1737=SKY\# 13794=
UBV M 13339=GCRV 5014=UBV 7255=LEE 197=HIC 36623=HIP 36623=
C* 779={\em IRAS} 07288+2436
\vspace*{-0.3cm}\\ 

\noindent
s07=WRAY 16$-$51=PK 271+03 1={\em IRAS} 09316$-$4621
\vspace*{-0.3cm}\\ 

\noindent
s08=Hen 3$-$653=SS73 30=PK 323$-$08 1=WRAY 15$-$807={\em IRAS} 11232$-$5940
\vspace*{-0.3cm}\\ 

\noindent
s09=NSV 05572=Ross 234=G 241$-$60=GD 407=LTT 16957
\vspace*{-0.3cm}\\ 

\noindent
s10=AE Cir=HV 5112
\vspace*{-0.3cm}\\ 

\noindent
s11=V748 Cen=CD$-$32 10517=CSV 2229=SON 5003=[OM87] 145632.12$-$331309.6
\vspace*{-0.3cm}\\ 

\noindent
s12=V345 Nor=HV 8827=NSV 07429=CSV 2543
\vspace*{-0.3cm}\\ 

\noindent
s13=V934 Her=HD 154791=BD+24 3121=3A 1703+241=SKY\# 30847=AG+24 1704=
YZ 24 5850=1E 1704.4+2402=2A 1704+241=2E 1704.4+2402=AT 1700+239=
AGKR 15254=PPM 105555=1H 1706+241=2E 3831=HIP 83714=SAO 84844=DO 15788=
4U 1700+24=HIC 83714=1ES 1704+24.0={\em IRAS} 17044+2402
\vspace*{-0.3cm}\\ 

\noindent
s14=Hen 3$-$1383=SS73 82=WRAY 15$-$1680
\vspace*{-0.3cm}\\ 

\noindent
s16=WSTB 19W032=PK 359+01 3=ESO 455$-$139=SCM 152=PN G359.2+01.2=
{\em IRAS} 17358$-$2854
\vspace*{-0.3cm}\\ 

\noindent
s17=WRAY 16$-$294=PK 002+02 1
\vspace*{-0.3cm}\\ 

\noindent
s18=AS 241=PN H 2$-$19=WRAY 16$-$302=PK 351$-$04 1=MHA 363$-$30=PN Sa 3$-$70=ESO 334$-$7=
PN VV' 238=Haro 3$-$19={\em IRAS} 17415$-$3816
\vspace*{-0.3cm}\\ 

\noindent
s20=V618 Sgr=HV 7203=AN 313.1933
\vspace*{-0.3cm}\\ 

\noindent
s21=AS 280=SS73 136=PK 358$-$06 1=Hen 2$-$354=WRAY 16$-$375=MHA 304$-$17=ESO 394$-$32
\vspace*{-0.3cm}\\ 

\noindent
s22=AS 288=PN H 2$-$43=PK 003$-$04 9=ESO 456$-$75=PN G003.4$-$04.8=MHA 304$-$119=
WRAY 16$-$388=PN VV' 367=Hen 2$-$366=Haro 3$-$43={\em IRAS} 18095$-$2820
\vspace*{-0.3cm}\\ 

\noindent
s23=Hen 2$-$379=SS73 150=PN M 1$-$44=PK 004$-$04 2=PN VV 165=WRAY 16$-$399=SCM 198=
ESO 522$-$11=PN G004.9$-$04.9=PN VV' 380={\em IRAS} 18131$-$2705
\vspace*{-0.3cm}\\ 

\noindent
s24=V335 Vul=AS 356=MHA 215$-$33=Case 452=LD 120=GSC 02128$-$00676=
C* 2728=CGCS 4253={\em IRAS} 19211+2421
\vspace*{-0.3cm}\\ 

\noindent
s25=V850 Aql=PN K 4$-$26=PK 037$-$06 2=CSV 4646=PN ARO 320={\em IRAS} 19210+0032
\vspace*{-0.3cm}\\ 

\noindent
s26=Hen 2$-$442=PN M 4$-$16=PK 061+02 1=PN G061.8+02.1=PN ARO 157
\vspace*{-0.3cm}\\ 

\noindent
s28=V627 Cas=AS 501=RAFGL 2999=MHA 73$-$59=HBC 316={\em IRAS} 22556+5833\\

\end{document}